\documentclass[runningheads]{llncs} 
\usepackage[T1]{fontenc}
\usepackage{graphicx}
\usepackage{amsmath}
\usepackage{amsfonts}
\usepackage{siunitx}
\usepackage{xcolor} 
\usepackage{colortbl}
\usepackage{url}
\usepackage{booktabs}
\usepackage[normalem]{ulem}
\usepackage{soul}
\newcommand{\DefineAuthor}[2]{%
  \expandafter\newcommand\csname #1\endcsname[1]{\textcolor{#2}{##1}}%
  \expandafter\newcommand\csname #1note\endcsname[1]{\textcolor{#2}{\hl{\textbf{\{#1:} \textit{##1}\textbf{\}}}}}%
  \expandafter\newcommand\csname #1todo\endcsname[1]{\textcolor{#2}{\hl{\textbf{\{#1.TODO:} \textit{##1}\textbf{\}}}}}%
  \expandafter\newcommand\csname #1del\endcsname[1]{\textcolor{#2}{\sout{##1}}}%
  \expandafter\newcommand\csname #1change\endcsname[2]{\textcolor{red}{\sout{##1}} \textcolor{#2}{##2}}%
}
\DefineAuthor{tk}{blue}    
\definecolor{darkgreen}{rgb}{0.0, 0.5, 0.0}

\begin{document}
\title{Benchmarking Machine Learning Approaches for Polarization Mapping in Ferroelectrics Using 4D-STEM}

\titlerunning{Polarization Mapping in Ferroelectrics Using 4D-STEM}
%
%
\author{Matej Martinc\orcidID{0000-0002-7384-8112
} \and
Goran Dražič\orcidID{0000-0001-7809-8050} \and
Anton Kokalj\orcidID{0000-0001-7237-0041} \and
Katarina Žiberna\orcidID{0009-0005-1613-8789} \and
Janina Roknić\orcidID{0009-0004-8118-1823} \and
Matic Poberžnik\orcidID{0000-0002-4866-4346} \and
Sašo Džeroski\orcidID{0000-0003-2363-712X} \and
Andreja Benčan Golob\orcidID{0000-0002-2116-3779}}
\authorrunning{Martinc et al.}
%
\institute{Jožef Stefan Institute, Jamova 39, Ljubljana, Slovenia\\
\email{\{name\}.\{surname\}@ijs.si}}
\maketitle              
\begin{abstract} Four-dimensional scanning transmission electron microscopy (4D-STEM) provides rich, atomic-scale insights into materials structures. However, extracting specific physical properties — such as polarization directions essential for understanding functional properties of ferroelectrics — remains a significant challenge. In this study, we systematically benchmark multiple machine learning models, namely ResNet, VGG, a custom convolutional neural network, and PCA-informed k-Nearest Neighbors, to automate the detection of polarization directions from 4D-STEM diffraction patterns in ferroelectric potassium sodium niobate. While models trained on synthetic data achieve high accuracy on idealized synthetic diffraction patterns of equivalent thickness, the domain gap between simulation and experiment remains a critical barrier to real-world deployment. In this context, a custom made prototype representation training regime and PCA-based methods, combined with data augmentation and filtering, can better bridge this gap. Error analysis reveals periodic missclassification patterns, indicating that not all diffraction patterns carry enough information for a successful classification. Additionally, our qualitative analysis demonstrates that irregularities in the model's prediction patterns correlate with defects in the crystal structure, suggesting that supervised models could be used for detecting structural defects. These findings guide the development of robust, transferable machine learning tools for electron microscopy analysis.


\end{abstract}

\section{Introduction}

Four-dimensional scanning transmission electron microscopy (4D-STEM) has emerged as a pivotal technique in materials science, offering unprecedented capabilities for characterizing materials down to atomic scale \cite{martis2023imaging,10.1017/S1431927619000497,ranieri2024assessing}. This advanced form of electron microscopy involves scanning a convergent electron beam across a two-dimensional region of a sample and, at each scan point, capturing the full two-dimensional diffraction pattern produced by the interaction of the electron beam with the specimen. The resulting dataset is four-dimensional, consisting of two spatial dimensions and two reciprocal space dimensions, providing a rich source of information about the local structural and electronic properties of materials. This technique facilitates a range of innovative imaging modalities, including crystal orientation mapping and phase contrast imaging, which are crucial for understanding the fundamental relationships between a material's structure and its functional properties \cite{sadri2025unsupervised}. The development of high-speed pixelated electron detectors, coupled with significant advancements in computational power, has been instrumental in the widespread adoption and application of 4D-STEM in materials research.

A critical aspect of characterizing ferroelectric materials, such as (K$_{0.5}$Na$_{0.5}$) NbO$_{3}$ (KNN) with its perovskite structure, is the determination of polarization direction and magnitude. KNN is a prominent and extensively researched environmentally friendly alternative to lead-based systems. Spontaneous polarization in these materials can lead to the formation of intricate domain structures at the nanoscale, which significantly influence their macroscopic properties and their suitability for various technological applications \cite{zhang2021lead}. The ability to accurately map polarization direction is thus important for advancing materials science. Traditional methods for analyzing 4D-STEM data often rely on manual inspection or rigid algorithms that require extensive prior knowledge. The features extracted from 4D-STEM data that are most relevant for polarization detection often include the asymmetry observed in diffraction patterns  \cite{zhu2024structural}. This asymmetry arises due to the polar nature of the material’s structure. The intensity distributions within the diffraction disks themselves can provide crucial information about crystal orientation, strain, and potential defects, which can all be indirectly linked to the polarization state of the material \cite{yuan2021machine}. The Center of Mass (CoM) of the diffraction pattern is another important feature, as it is directly related to the local electric fields present in the sample, which are intrinsically connected to polarization in ferroelectric materials \cite{cao2021new}.

Traditional methods are time-consuming and may struggle to identify anomalous features or subtle polarization shifts \cite{Bencan_NC12,bruefach2022analysis,Condurache_NL23,Rojac_NM16}. While machine learning (ML) offers a promising avenue for automation, the optimal strategy for deploying these algorithms, specifically regarding architecture selection and the transferability of models trained on simulated data to experimental settings, remains an open research question. ML methods, especially deep learning, have been applied for various analyses of electron microscopy data, including the 4D-STEM datasets \cite{10.1093/micmic/ozad067.1015,sadri2025unsupervised,cao2021new,roccapriore2022automated}. Among the prevalent supervised learning methods are Convolutional Neural Networks (CNNs), which have been applied for tasks like estimating sample thickness, identifying complex nanostructures, and mapping different phases \cite{10.1093/micmic/ozad067.1015,nordahl2024exploring,sadri2025unsupervised,hardy2025polarization,yuan2021machine,zhu2024structural}. On the other hand, unsupervised learning algorithms such as Non-negative Matrix Factorization (NMF) and Principal Component Analysis (PCA) play a crucial role in reducing the dimensionality of large 4D-STEM datasets, denoising data, and identifying underlying structural features without requiring prior knowledge or labeled data \cite{bruefach2022analysis,bruefach2023robust,kimoto2024unsupervised,10.1093/micmic/ozad067.1015}. 

While recent studies have utilized neural networks for polarization mapping \cite{yuan2021machine,hardy2025polarization} and domain separation \cite{ludacka2024imaging}, this work differentiates itself from related literature by moving beyond the demonstration of a single successful model on a specific dataset. Instead, our research focuses on the comparative analysis of supervised strategies and the explicit evaluation of the synthetic-to-experimental domain gap. We provide a detailed examination of how different architectures and learning objectives handle the complexity and noise inherent in real experimental data when ground truth is derived from idealized simulations. We evaluate a supervised learning framework for classifying polarization directions in electron diffraction patterns across several distinct scenarios. In the first phase, supervised polarization classification models are trained on 4D-STEM simulations of a 20 nm thick crystal specimen (2×2×50 unit cells) using a randomized 80/10/10 split for training, validation/prototyping, and initial testing. To generate the dataset, we simulated 8 different  4D-STEM models, each with a different polarization direction. In the second phase, the robustness and transferability of the trained supervised models are tested against varying specimen geometries, including a laterally larger specimen (20~nm thick, 6×6×50 unit cells) and a specimen that is both laterally larger and thicker (50~nm thick, 6×6×125 unit cells). In the third phase, the trained supervised models are tested on experimental diffraction data, for which we had previously manually identified the polarization direction based on HAADF image analysis. Finally, we test the supervised models’ ability to identify a position of  defect in  simulated 4D-STEM models which, according to the simulations, manifests as a local change in polarization.

The specific contributions of this study are:
\begin{itemize}
\item A systematic benchmark of pretrained convolutional networks (ResNet \cite{he2016deep} and VGG \cite{simonyan2014very}), a custom convolutional network, and a k-Nearest Neighbors (k-NN) approach utilizing Principal Component Analysis (PCA) feature extraction for polarization classification.
\item Investigation and comparison of distinct training paradigms and learning objectives, namely standard classification (predicting a label), regression (predicting the polarization angle in radians), and a prototype representation approach (learning a representative embedding for each polarization direction).
\item A critical analysis of the domain gap in 4D-STEM, demonstrating that while standard deep learning models excel in synthetic environments, specialized approaches like prototype learning and PCA-based approach are better at bridging the gap to experimental data. Additionally, we show that the domain gap can be partially overcome by  augmentation and filtering employed on the synthetic train dataset. 
\item A comprehensive error analysis exploring the inner workings of our best models and how periodic patterns in model missclassifications correlate with the properties and structure of 4D-STEM datasets. Finally, we also test the utility of supervised models trained on synthetic patterns for the detection of anomalies and show that the missclassification patterns can be potentially used for detection of structural defects.
\end{itemize}

\section{Methods}

This study evaluates supervised learning strategies for polarization classification in 4D-STEM. We systematically benchmark different architectures (ResNet, VGG, Custom CNN, k-NN) and training paradigms (classification, regression, prototype learning) to determine the spontaneous polarization direction in a perovskite KNN crystal.

Two evaluation scenarios are considered. In the synthetic-to-synthetic setting, models trained on synthetic data are evaluated on additional synthetic datasets. In the synthetic-to-experimental setting, we assess how well models trained on synthetic diffraction patterns generalize to experimental 4D-STEM data consisting of $128 \times 128 = 16\,384$ diffraction patterns.

The methodology consists of three main stages. First, we generate synthetic 4D-STEM datasets. Second, we introduce preprocessing techniques designed to reduce the domain gap between synthetic training data and experimental test data. Third, we evaluate multiple models and training paradigms on the task of polarization classification.

\subsection{Synthetic Dataset Generation}
\label{sec:synth-dataset}

The studies investigating the application of machine learning for polarization- direction detection in 4D-STEM utilize a variety of datasets, both experimental and simulated \cite{bruefach2022analysis}. While experimental datasets are essential for validating model performance on real-world data and for discovering unanticipated phenomena, acquiring sufficiently large and well-annotated experimental 4D-STEM datasets is often extremely challenging or even infeasible, especially when specific polarization directions are required. To address this limitation, previous studies have proposed training models on simulated data \cite{oxley2020deep}, leveraging the ability to learn complex structural relationships in a controlled environment and then applying domain adaptation and over-fitting prevention techniques to transfer these models to experimental data. We adopt this approach in our work.

\begin{figure}[htb]
  \centering
  \includegraphics[width=0.95\textwidth]{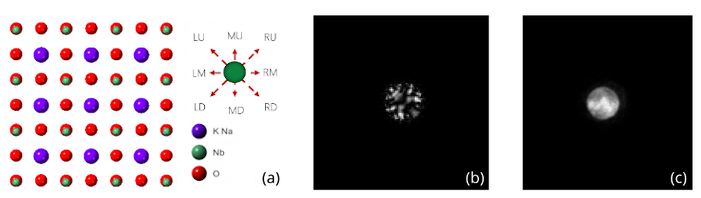}
  \caption{(a) Crystal structure of orthorhombic KNN viewed along the $[100]_{\rm pc}$ zone axis with a graph showing 8 possible Nb displacement directions, and examples of the corresponding (b) simulated and (c) experimental diffraction patterns.}
  \label{img-knn} 
\end{figure}

For training the models, we generate simulated 4D-STEM diffraction patterns for lead-free KNN ferroelectric (Figure~\ref{img-knn}a, ICSD \#186310), based on $2\times2\times50$ unit cell model with known Nb displacements using QSTEM simulation \cite{koch2002determination}, performed by using 24 mrad convergent semi-angle, 20 nm thickness, real-image size of $128 \times 128$ pixels, and reciprocal-image (diffraction pattern) size of $256 \times 256$ pixels. KNN’s orthorhombic symmetry allows for 12 possible off-center displacement directions of the Nb atoms, correlated to the polarization vector direction \cite{Chen_NC13}. When projected onto 2D, these displacements can be seen in 8 distinct directions along the $[100]_{\rm pc}$ zone axis, namely left-up (LU), right-up (RU), right-middle (RM), left-down (LD), right-down (RD), middle-up (MU), right-middle (RM), and left-middle (LM) (Figure~\ref{img-knn}a). The final training dataset contains 16384 images per class (i.e., direction), corresponding to a $128 \times 128$ grid of synthetic diffraction patterns (examples of synthetic and experimental diffraction patterns are shown in Figure~\ref{img-knn}b,c).

To verify the robustness of the method during the testing phase, we generate additional 4D-STEM datasets with the following dimensions: $4\times4\times50$ for the LD (20 nm thick), $6\times6\times50$ unit cells for both the LU and RU (each 20 nm thick) and $6\times6\times125$ unit cells (50 nm thick) for LU and RD configurations. Furthermore, to test defect detection, we created a synthetic test set by simulating an oxygen vacancy. This was done by removing one octahedrally coordinated O ion per 640-atom supercell in the RD configuration and relaxing the resulting structure using DFT calculations performed with Quantum ESPRESSO \cite{Giannozzi_JPCM21,Giannozzi_JPCM29} within the PBE exchange--correlation functional \cite{Perdew_PRL77} and PAW pseudopotentials \cite{Blochl_PRB50} from the pslibrary \cite{DalCorso_CMS95,PAW}. A plane-wave basis set with a kinetic-energy cutoff of 50~Ry was used, and electron correlation effects were treated within the DFT+U framework \cite{Anisimov_PRB44,Cococcioni_PRB71} with the Hubbard $U$ parameter determined using the \texttt{hp.x} module \cite{Timrov_PRB98}, yielding $U=2.4$~eV.

\subsection{Preprocessing}
\label{sec:preprocessing}

As noted above, there is a big domain shift between simulated images on which the model is trained and the experimental images on which the model is applied. If no augmentation is applied, the model is learning on ``clean'' simulated features from the QSTEM simulations (e.g., exact pixel intensities, zero background) that might not exist in the experimental data. This causes the model to converge to a synthetic local minimum very different from the minimum in the experimental data, which creates arbitrary decision boundaries, leading to unstable and unreliable predictions. To prevent this to happen, we also introduce extensive preprocessing.

To isolate the diffraction signal and correct for systematic experimental drifts, we implemented a custom cropping strategy based on the CoM. First, a global average CoM was calculated for each dataset to quantify the systematic misalignment of the detector relative to the optical axis. A corrected cropping coordinate was then derived by subtracting the systematic global offset from the image center. This ensures that the $64 \times 64$ pixel crop remains centered on the diffraction disk while preserving the subtle local displacements indicative of ferroelectric polarization. More specifically, CoM is calculated in the following manner. Let \( I(x, y) \) be the intensity at pixel coordinate \( (x, y) \). Then the CoM coordinates \( (x_{\text{CoM}}, y_{\text{CoM}}) \) are given by: 
\begin{equation}
  \label{eq:CoM}
x_{\text{CoM}} = \frac{\sum\limits_{x, y} x I(x, y)}{\sum\limits_{x, y} I(x, y)}, \qquad
y_{\text{CoM}} = \frac{\sum\limits_{x, y} y I(x, y)}{\sum\limits_{x, y} I(x, y)}.  
\end{equation}
The displacement from the center of the pattern $(x_{\text{center}}, y_{\text{center}})$ can be calculated as:
\begin{equation}
  \label{eq:delta-x,y}
{\rm\Delta} x = x_{\text{CoM}} - x_{\text{center}}, \qquad
{\rm\Delta} y = y_{\text{CoM}} - y_{\text{center}}.
\end{equation}
By using ${\rm\Delta} x$ and ${\rm\Delta} y$, we define the electron beam shift magnitude ($M$) as:
\begin{equation}
  \label{eq:M}
  M = \sqrt{{\rm\Delta} x^2 + {\rm\Delta} y^2}.
\end{equation}

Following the cropping, each image in the TIFF format is converted to a tensor, and pixel values are normalized to values between 0 and 1. We conduct cropping and normalization on all the images. 

We apply additional preprocessing only on the training images (i.e., synthetic training diffraction patterns). To enhance the stability and convergence of the classifier, we implemented a filtering protocol for the synthetic training data based on the magnitude $M$ of Eq.~\eqref{eq:M}. In ferroelectric materials, unit cells with weak polarization displacements generate diffraction patterns that are nearly centrosymmetric. Such patterns exhibit high visual similarity across all directional classes, effectively acting as ambiguous labels that hinder the learning of discriminative features. By restricting the training set to simulated patterns with above-average beam shift magnitudes, we ensure that the network is trained exclusively on exemplars with visible polarization. By doing this, we want to prevent the model from overfitting to negligible pixel variations in quasi-symmetric patterns and create sharper decision boundaries in the latent space. Consequently, the model learns robust directional features that generalize better to experimental data, where signal-to-noise ratios are inherently lower. More specifically, we only train on simulated images with above average electron beam shift magnitude ($M$), which gives us a slightly unbalanced polarization distribution for the synthetic training dataset ($D_t$), containing roughly 7000 examples per class. The exact training dataset class distribution after filtering is the following: 
$$
{\scriptstyle D_t = \{{\rm MD}: 7244,\ {\rm LM:} 7243,\ {\rm LD:} 7118,\ {\rm LU:} 7077,\ {\rm RM:} 7020,\ {\rm MU:} 7020,\ {\rm RD:} 6963,\ {\rm RU:} 6946\}}.
$$
Only during training, we apply a randomized data augmentation pipeline using the Albumentations\footnote{\url{https://pypi.org/project/albumentations/}} library to improve model robustness against experimental variations. To this end, we employ the following augmentation operations:
\begin{itemize}
    \item \textbf{Adaptive Gaussian Smoothing}: We apply a Gaussian filter with a randomized variance $\sigma \in [2,5]$. This kernel size is specifically chosen to be mild, ensuring that essential structural features—such as CoM shifts in diffraction peaks—remain preserved while simulating slight loss of focus.
    \item \textbf{Brightness and Contrast Adjustment}: Random fluctuations in intensity and local contrast (limited to $\pm15\%$) are applied to simulate variations in exposure levels and detector sensitivity.
    \item \textbf{Noise Injection}: Gaussian noise is added to mimic experimental detector read noise and electron shot noise.
\end{itemize}
Each augmentation is applied to the training images with a predefined probability: Adaptive Gaussian Smoothing is applied to the train images with a 0.5 probability, Brightness and Contrast Adjustment with a 0.3 probability and Noise Injection with a 0.2 probability.

\subsection{Polarization Classification}

We train our models on the generated synthetic data. More specifically,  80\% of images from $D_t$ are used for training, while we leave random 10\% of synthetic data aside for validation and prototype representation construction (as explained below), and 10\% for model testing on synthetic data. 

We use three distinct neural encoders to derive latent feature representations. We use two pretrained models, ResNet \cite{he2016deep} and VGG \cite{simonyan2014very}, both pretrained on ImageNet \cite{deng2009imagenet}. We modify these two models to support 1-channel input images by creating an input embedding containing averaged original RGB weights from the original model. Since there is a substantial domain mismatch between images in ImageNet and (simulated and experimental) 4D-STEM data, which might prevent effective transfer of pre-learned features, and since we have a sufficient amount of training data, we additionally train a convolutional (CONV) model from scratch. We opted for a shallow model with just three convolutional layers to prevent overfitting. Let $\mathbf{X} \in \mathbb{R}^{1 \times H \times W}$ be the input (single-channel image). Our custom CONV encoder is defined as:
\begin{align*}
\mathbf{X}_0 &= \mathbf{X} \in \mathbb{R}^{1 \times H \times W} \\
\mathbf{X}_1 &= \sigma\left( \text{Conv}_{1 \rightarrow 16, 3, 1}(\mathbf{X}_0) \right) \\
\mathbf{X}_2 &= \sigma\left( \text{Conv}_{16 \rightarrow 32, 3, 1}(\mathbf{X}_1) \right) \\
\mathbf{X}_3 &= \text{AvgPool}_2(\mathbf{X}_2) \\
\mathbf{X}_4 &= \sigma\left( \text{Conv}_{32 \rightarrow 64, 3, 1}(\mathbf{X}_3) \right) \\
\mathbf{X}_5 &= \text{AvgPool}_2(\mathbf{X}_4)
\end{align*}
Each \texttt{Conv2d(in\_channels, out\_channels, kernel\_size=3, stride=1)} is denoted as $\text{Conv}_{\text{in} \rightarrow \text{out}, 3, 1}$. Each ReLU activation is written as $\sigma(\cdot)$ and \texttt{AvgPool2d(2)} is written as $\text{AvgPool}_2(\cdot)$ with stride equal to the kernel size.

We test four distinct approaches to derive the polarization direction in the structure.

\subsubsection{Classification Approach:}

By adding a dense classification layer (with output size equal to the number of polarization classes) on top of the encoder, we train the model to produce a correct polarization label for each diffraction pattern in a structure with a specific polarization. We use a standard cross entropy loss to train the model to predict probabilities for each of the 8 polarization classes for each diffraction pattern. During inference, the class with the maximum probability is chosen as a correct class for each diffraction pattern. 

\subsubsection{Regression Approach:}

While classification treats polarization directions as independent categories, the physical polarization directions are continuous and spatially related (e.g., $0^\circ$ is geometrically closer to $45^\circ$ than to $180^\circ$). To leverage this geometric relationship, we employ a \textbf{regression} approach. This model utilizes the same encoder architecture as the classification approach but is equipped with a specific regression head consisting of a linear layer with 2 outputs followed by a Tanh activation function. This head is trained to predict a 2D unit vector $\mathbf{v} = (v_x, v_y)$ representing the polarization direction, where the values are constrained to the interval $[-1, 1]$. The ground truth class labels are mapped to continuous target vectors on the unit circle using the following transformation, where $k$ is the class index:
\begin{equation}
  \label{eq:theta_k}
  \theta_k = k \times \frac{\pi}{4}, \quad \mathbf{v}_{\rm target} = (\cos \theta_k, \sin \theta_k).
\end{equation}
We utilize the Mean Squared Error (MSE) loss to train the model:
\begin{equation}
  \label{eq:MSE}
\mathcal{L}_{\rm MSE} = \frac{1}{N} \sum_{i=1}^{N} \|\mathbf{v}_{\rm pred}^{(i)} - \mathbf{v}_{\rm target}^{(i)}\|^2.  
\end{equation}
During inference, the predicted vector is converted back to a discrete class index by computing the angle $\phi = \text{atan2}(v_y, v_x)$. We apply a bucketization strategy where the angle is shifted by $22.5^\circ$ ($\frac{\pi}{8}$ radians) to center the bins around the primary polarization directions, effectively discretizing the continuous prediction into the nearest of the 8 classes.

\subsubsection{Prototype Representation Approach:}

The third approach is based on two important insights/observations specific to 4D-STEM. First, in 4D-STEM, a full diffraction pattern is collected at each scan point, meaning that each pattern carries some global structural information on polarization-induced asymmetry of an entire structure. Second, the experimental dataset contains a lot of noise that makes predictions on the level of single diffraction patterns unreliable. Additionally, this approach also considers geometric relationships between classes, same as the regression approach.

We attach a dense layer to the encoder (output size $D = 128$) to generate an embedding vector $\mathbf{x}_i$ for each input. First, we compute the prototypes $\mathbf{c}_k$ for each class $k$ in the current batch by averaging the embeddings of samples belonging to that class:
\begin{equation}
  \label{eq:c_k}
  \mathbf{c}_k = \frac{1}{|\mathcal{I}_k|} \sum_{j \in \mathcal{I}_k} \mathbf{x}_j,  
\end{equation}
where $\mathcal{I}_k$ is the set of indices for samples with label $k$.

Unlike standard prototypical networks that enforce hard assignment, we construct a soft target distribution that reflects the circular nature of the polarization directions. For a sample with true class label $y_i$, the circular distance to any other class $k$ is defined as:
\begin{equation}
  \label{eq:d_circ}
  d_{\rm circ}(y_i, k) = \min(|y_i - k|, C - |y_i - k|),
\end{equation}

where $C=8$ is the number of classes. This ensures that class 0 and class 7 are treated as neighbors. We then generate target probabilities $P_{i,k}^{\rm target}$ using a Gaussian kernel controlled by a parameter $\sigma$ (set to 1.25), which allows the model to be more forgiving of predictions near the true angle:
\begin{equation}
  \label{eq:w_i,k}
  w_{i,k} = \exp\left(-\frac{d_{\rm circ}(y_i, k)^2}{2\sigma^2}\right), \quad P_{i,k}^{\rm target} = \frac{w_{i,k}}{\sum_{j=1}^C w_{i,j}}.  
\end{equation}
The network's predicted log-probabilities are derived from the negative Euclidean distances between the embedding $\mathbf{x}_i$ and the prototypes:
\begin{equation}
  \label{eq:logP_i,k}
\log P_{i,k}^{\rm pred} = \log \left( \frac{\exp(-\|\mathbf{x}_i - \mathbf{c}_k\|)}{\sum_{j=1}^C \exp(-\|\mathbf{x}_i - \mathbf{c}_j\|)} \right).  
\end{equation}
Finally, the model is trained by minimizing the Kullback-Leibler (KL) Divergence between the soft target distribution and the predicted distribution:
\begin{equation}
  \label{eq:L_circ}
\mathcal{L}_{\rm circ} = \frac{1}{N} \sum_{i=1}^N \sum_{k=1}^C P_{i,k}^{\rm target} \log \left( \frac{P_{i,k}^{\rm target}}{P_{i,k}^{\rm pred}} \right).  
\end{equation}

\begin{figure}[ht]
\includegraphics[width=0.9\textwidth]{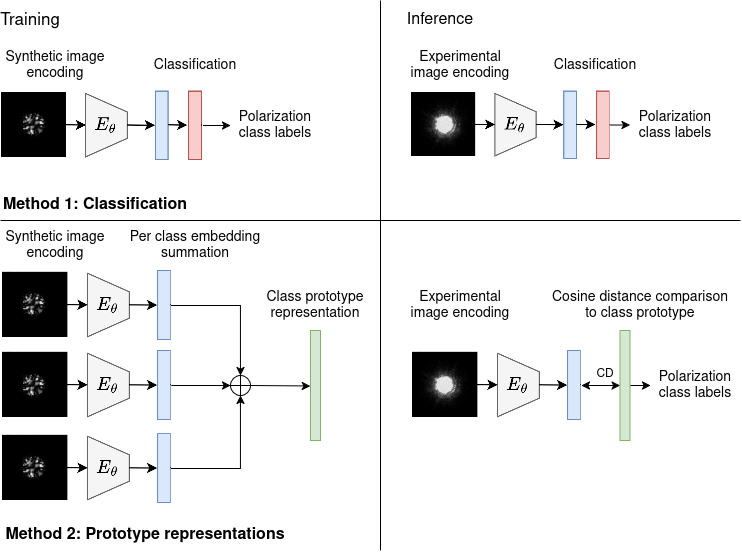} 
\caption{Methodologies used. The upper part shows standard classification approaches and regression with mapping to classes, while the lower part shows the prototype-representation-based approach.}
  \label{img-methodology}
\end{figure}

During inference, we compute fixed prototypes using the synthetic validation set  (i.e., 10\% of $D_t$). For each new diffraction pattern, we assign the label of the prototype with the highest cosine similarity to the pattern's embedding. The difference between the prototype representation approach and the other two neural approaches is described in Figure \ref{img-methodology}.

\subsubsection{PCA + k-NN:}

To determine if deep feature extraction is strictly necessary, we implement a classic machine learning pipeline utilizing Principal Component Analysis (PCA) and k-Nearest Neighbors (k-NN). First, the diffraction patterns are flattened and scaled. We apply PCA to reduce the dimensionality of the data, retaining the top 20 principal components ($n_{\rm components}=20$) which capture the most significant variance in the diffraction signal. These reduced feature vectors $\mathbf{z} \in \mathbb{R}^{20}$ are then used to train a k-NN classifier with $k=5$ neighbors. During inference, the label is determined by a majority vote among the 5 nearest neighbors in the projected PCA space.

\subsection{Experimental Setting and Evaluation}

All models are trained on 80\% of the generated synthetic train datasets described in Section \ref{sec:synth-dataset}. For neural models, 10\% of diffraction patterns per class are randomly set aside for testing and 10\% for validation and prototype representation. Since the PCA+k-NN approach does not require a validation set, here the model is trained on 90\% of data and 10\% of diffraction patterns per class are randomly set aside for testing. We use batch size of 128 and a learning rate of $2\times10^{-4}$ for all neural models and for all three approaches: the classification approach, the regression approach and the approach based on prototype representations. All neural models are trained up to 20 epochs and we employ early stopping if the validation loss does not improve for 5 epochs. To improve the robustness of our experimental setting, we train five different model versions using five random seeds per each model and approach.

We perform an ablation study to determine how different preprocessing techniques affect the performance. Specifically, we test three preprocessing configurations:
\begin{itemize}
    \item \textbf{No augmentation and no filtering}: The training data are used without augmentation (Gaussian smoothing, brightness and contrast adjustment, or noise injection) and without filtering based on the magnitude of the polarization.
    \item \textbf{Augmentation and no filtering}: We apply augmentation of the training dataset but we do not conduct training dataset filtering based on the magnitude of the polarization.
    \item \textbf{Augmentation and filtering}: We apply both augmentation and filtering on the training dataset.
\end{itemize}

To test all models and approaches, we use a set of synthetic test datasets described in Section  \ref{sec:synth-dataset} and a set of experimental 4D-STEM datasets. The characteristics of the experimental datasets, such as size, dimensionality, and the information content of the diffraction patterns, are influenced by various experimental parameters, including the convergence angle of the electron beam, the beam current, and the dwell time at each scan point \cite{ranieri2024assessing}. Different STEM operating modes and detector configurations can also affect the nature of the acquired data \cite{10.1017/S1431927619000497}. More specifically, the Cs-corrected scanning transmission electron microscope ARM200CF (Jeol, ARM) was used to acquire 4D-STEM from a KNN crystal of ~ 20 nm thickness using the pixelated detector Merlin (Quantum detectors, UK) at a convergence angle of 24 mrad. We obtained an experimental test set corresponding to a $128 \times 128$ grid of experimental diffraction patterns. Each pattern was a $256 \times 256$ pixel grayscale TIFF image. The polarization vector direction, estimated manually from quantitative analysis of Nb displacements relative to the alkali sublattice in STEM images (procedure is described in ref.~\cite{condurache2021atomically}), was used to label each image.

\section{Results}

\subsection{Polarization Classification}

The results are presented across three distinct training configurations: no augmentation and no filtering (Table~\ref{tab:results_detailed_1}), augmentation and no filtering (Table~\ref{tab:results_detailed_2}), and augmentation and filtering (Table~\ref{tab:results_detailed_3}). These configurations define how the synthetic training data was processed, specifically concerning data augmentation (applying Gaussian Smoothing, Brightness and Contrast Adjustment, and Noise Injection) and filtering (selecting training samples based on the magnitude of the polarization). The performance of each model is assessed using two main metrics:

\begin{itemize}
    \item \textbf{Accuracy (Top Row):} This measures the overall percentage of correctly classified diffraction patterns and is the primary performance indicator. The result is presented as the mean ± standard deviation across five random seeds. The highest mean accuracy in each row is bolded. Note that for 4E-RM/RD and 5E-RM/RD experimental datasets, where two solutions (polarization directions) are possible, we count the prediction of either of the two solutions as correct.
    \item \textbf{Majority Class Prediction (Bottom Row):} This indicates the polarization class that the model most frequently predicted for all diffraction patterns within a given test structure. Since the majority class represents the overall polarization direction for an entire structure, this metric reflects the model's ability to capture the macro-level domain property. If the majority class is colored green, this means that the predicted majority class matches the true (ground truth) majority class. Otherwise, the class is colored red.
\end{itemize}

\begin{table*}[ht]
  \caption{Accuracy (top row) and majority class prediction (bottom row) for the \textbf{no augmentation and no filtering} configuration. The highest accuracy in each row is \textbf{bolded}. \textcolor{darkgreen}{Green} indicates the predicted majority matches the ground truth, while \textcolor{red}{red} indicates a mismatch.}
\label{tab:results_detailed_1}
\centering
\resizebox{\textwidth}{!}{
\setlength{\tabcolsep}{2pt}
\begin{tabular}{lcccccccccc}
\toprule
Dataset & Conv (Reg) & Conv (Cls) & Conv (Proto) & VGG (Reg) & VGG (Cls) & VGG (Proto) & ResNet (Reg) & ResNet (Cls) & ResNet (Proto) & PCA \\
\midrule
\textbf{Synthetic test} & $100.0 \pm 0.0$ & $\mathbf{100.0} \pm \mathbf{0.0}$ & $99.6 \pm 0.5$ & $78.0 \pm 26.9$ & $70.6 \pm 36.0$ & $84.2 \pm 13.1$ & $99.5 \pm 0.2$ & $99.8 \pm 0.3$ & $95.4 \pm 0.8$ & $95.1 \pm 0.2$ \\[0.4em]
\textbf{1S-LU-6x6-20nm} & $0.0 \pm 0.0$ & $43.2 \pm 2.7$ & $47.5 \pm 3.0$ & $0.0 \pm 0.0$ & $41.8 \pm 28.8$ & $54.3 \pm 12.5$ & $0.0 \pm 0.0$ & $\mathbf{63.1} \pm \mathbf{2.2}$ & $50.6 \pm 3.9$ & $24.8 \pm 0.0$ \\
\textit{Maj. Class} & \textcolor{red}{RU} & \textcolor{darkgreen}{LU} & \textcolor{darkgreen}{LU} & \textcolor{red}{RU} & \textcolor{darkgreen}{LU} & \textcolor{darkgreen}{LU} & \textcolor{red}{RU} & \textcolor{darkgreen}{LU} & \textcolor{darkgreen}{LU} & \textcolor{darkgreen}{LU} \\[0.2em]
\textbf{3S-RM-2x2-20nm} & $\mathbf{100.0} \pm \mathbf{0.0}$ & $\mathbf{100.0} \pm \mathbf{0.0}$ & $99.9 \pm 0.0$ & $99.9 \pm 0.2$ & $70.6 \pm 39.6$ & $76.2 \pm 17.9$ & $100.0 \pm 0.0$ & $99.8 \pm 0.3$ & $95.5 \pm 2.0$ & $97.9 \pm 0.1$ \\
\textit{Maj. Class} & \textcolor{darkgreen}{RM} & \textcolor{darkgreen}{RM} & \textcolor{darkgreen}{RM} & \textcolor{darkgreen}{RM} & \textcolor{darkgreen}{RM} & \textcolor{darkgreen}{RM} & \textcolor{darkgreen}{RM} & \textcolor{darkgreen}{RM} & \textcolor{darkgreen}{RM} & \textcolor{darkgreen}{RM} \\[0.2em]
\textbf{5S-RU-6x6-20nm} & $41.8 \pm 7.5$ & $96.5 \pm 1.1$ & $95.5 \pm 0.7$ & $43.9 \pm 28.5$ & $63.6 \pm 36.4$ & $77.0 \pm 16.5$ & $37.9 \pm 26.8$ & $\mathbf{98.4} \pm \mathbf{0.4}$ & $91.0 \pm 2.2$ & $52.0 \pm 0.2$ \\[0.2em]
\textit{Maj. Class} & \textcolor{red}{RM} & \textcolor{darkgreen}{RU} & \textcolor{darkgreen}{RU} & \textcolor{red}{RM} & \textcolor{darkgreen}{RU} & \textcolor{darkgreen}{RU} & \textcolor{red}{RM} & \textcolor{darkgreen}{RU} & \textcolor{darkgreen}{RU} & \textcolor{darkgreen}{RU} \\[0.2em]
\textbf{6S-LD-4x4-20nm} & $0.0 \pm 0.0$ & $97.0 \pm 0.8$ & $\mathbf{98.1} \pm \mathbf{1.3}$ & $0.0 \pm 0.0$ & $82.9 \pm 21.5$ & $89.6 \pm 9.3$ & $0.0 \pm 0.0$ & $97.3 \pm 0.6$ & $91.0 \pm 1.9$ & $94.3 \pm 0.2$ \\[0.2em]
\textit{Maj. Class} & \textcolor{red}{RM} & \textcolor{darkgreen}{LD} & \textcolor{darkgreen}{LD} & \textcolor{red}{RU} & \textcolor{darkgreen}{LD} & \textcolor{darkgreen}{LD} & \textcolor{red}{RU} & \textcolor{darkgreen}{LD} & \textcolor{darkgreen}{LD} & \textcolor{darkgreen}{LD} \\[0.2em]
\textbf{7S-RD-2x2-20nm} & $0.0 \pm 0.0$ & $\mathbf{100.0} \pm \mathbf{0.0}$ & $99.8 \pm 0.4$ & $0.0 \pm 0.0$ & $60.3 \pm 48.6$ & $91.0 \pm 7.1$ & $0.0 \pm 0.0$ & $99.3 \pm 1.2$ & $96.9 \pm 1.8$ & $97.4 \pm 0.1$ \\[0.2em]
\textit{Maj. Class} & \textcolor{red}{RM} & \textcolor{darkgreen}{RD} & \textcolor{darkgreen}{RD} & \textcolor{red}{RM} & \textcolor{darkgreen}{RD} & \textcolor{darkgreen}{RD} & \textcolor{red}{RM} & \textcolor{darkgreen}{RD} & \textcolor{darkgreen}{RD} & \textcolor{darkgreen}{RD} \\[0.2em]
\textbf{2S-LU-6x6-50nm} & $0.0 \pm 0.0$ & $10.6 \pm 1.2$ & $\mathbf{12.7} \pm \mathbf{4.9}$ & $0.0 \pm 0.0$ & $2.0 \pm 3.9$ & $10.0 \pm 17.3$ & $0.0 \pm 0.0$ & $2.1 \pm 0.8$ & $7.6 \pm 1.1$ & $7.6 \pm 0.3$ \\[0.2em]
\textit{Maj. Class} & \textcolor{red}{RM} & \textcolor{red}{RM} & \textcolor{red}{RM} & \textcolor{red}{RM} & \textcolor{red}{MD} & \textcolor{red}{MD} & \textcolor{red}{RU} & \textcolor{red}{MU} & \textcolor{red}{RM} & \textcolor{red}{LD} \\[0.2em]
\textbf{4S-RD-6x6-50nm} & $0.0 \pm 0.0$ & $13.7 \pm 2.2$ & $12.1 \pm 4.3$ & $0.0 \pm 0.0$ & $0.6 \pm 1.1$ & $14.9 \pm 15.7$ & $0.0 \pm 0.0$ & $4.5 \pm 3.5$ & $13.4 \pm 4.1$ & $\mathbf{31.0} \pm \mathbf{0.3}$ \\[0.2em]
\textit{Maj. Class} & \textcolor{red}{RM} & \textcolor{red}{RM} & \textcolor{red}{RU} & \textcolor{red}{RM} & \textcolor{red}{RM} & \textcolor{red}{MD} & \textcolor{red}{RM} & \textcolor{red}{MU} & \textcolor{red}{RM} & \textcolor{darkgreen}{RD} \\[0.2em]
\textbf{1E-RM} & $25.4 \pm 37.5$ & $11.1 \pm 7.8$ & $19.4 \pm 27.8$ & $\mathbf{46.9} \pm \mathbf{40.5}$ & $20.0 \pm 40.0$ & $17.4 \pm 20.2$ & $38.0 \pm 46.6$ & $0.0 \pm 0.0$ & $1.1 \pm 2.2$ & $0.0 \pm 0.0$ \\[0.2em]
\textit{Maj. Class} & \textcolor{red}{RU} & \textcolor{red}{LD} & \textcolor{red}{LM} & \textcolor{red}{RU} & \textcolor{red}{LD} & \textcolor{red}{RD} & \textcolor{red}{RU} & \textcolor{red}{RU} & \textcolor{red}{LD} & \textcolor{red}{MD} \\[0.2em]
\textbf{2E-RM} & $25.4 \pm 37.5$ & $11.1 \pm 7.8$ & $19.4 \pm 27.8$ & $\mathbf{46.9} \pm \mathbf{40.5}$ & $20.0 \pm 40.0$ & $17.4 \pm 20.2$ & $38.0 \pm 46.6$ & $0.0 \pm 0.0$ & $1.1 \pm 2.2$ & $0.0 \pm 0.0$ \\[0.2em]
\textit{Maj. Class} & \textcolor{red}{RU} & \textcolor{red}{LD} & \textcolor{red}{LM} & \textcolor{red}{RU} & \textcolor{red}{LD} & \textcolor{red}{RD} & \textcolor{red}{RU} & \textcolor{red}{RU} & \textcolor{red}{LD} & \textcolor{red}{MD} \\[0.2em]
\textbf{3E-RM} & $24.4 \pm 37.1$ & $8.7 \pm 5.8$ & $17.3 \pm 25.9$ & $\mathbf{46.5} \pm \mathbf{41.9}$ & $20.0 \pm 40.0$ & $18.0 \pm 20.5$ & $37.5 \pm 46.0$ & $0.0 \pm 0.0$ & $1.2 \pm 2.5$ & $0.0 \pm 0.0$ \\[0.2em]
\textit{Maj. Class} & \textcolor{red}{RU} & \textcolor{red}{LD} & \textcolor{red}{LM} & \textcolor{red}{RU} & \textcolor{red}{LD} & \textcolor{red}{RD} & \textcolor{red}{RU} & \textcolor{red}{RU} & \textcolor{red}{LD} & \textcolor{red}{MD} \\[0.2em]
\textbf{4E-RM/RD} & $29.2 \pm 33.4$ & $7.4 \pm 5.3$ & $17.1 \pm 15.0$ & $\mathbf{40.2} \pm \mathbf{37.3}$ & $38.9 \pm 46.5$ & $4.4 \pm 7.7$ & $34.9 \pm 43.5$ & $7.7 \pm 10.5$ & $17.5 \pm 34.9$ & $0.0 \pm 0.0$ \\[0.2em]
\textit{Maj. Class} & \textcolor{red}{RU} & \textcolor{red}{LD} & \textcolor{red}{LM} & \textcolor{red}{RU} & \textcolor{darkgreen}{RD} & \textcolor{red}{LU} & \textcolor{red}{RU} & \textcolor{red}{RU} & \textcolor{red}{LD} & \textcolor{red}{LM} \\[0.2em]
\textbf{5E-RM/RD} & $31.4 \pm 37.7$ & $36.2 \pm 33.0$ & $9.0 \pm 9.1$ & $\mathbf{39.9} \pm \mathbf{36.0}$ & $35.4 \pm 43.9$ & $20.1 \pm 24.3$ & $36.8 \pm 45.4$ & $7.6 \pm 11.2$ & $20.0 \pm 39.9$ & $0.0 \pm 0.0$ \\[0.2em]
\textit{Maj. Class} & \textcolor{red}{RU} & \textcolor{darkgreen}{RD} & \textcolor{red}{LD} & \textcolor{red}{RU} & \textcolor{red}{LD} & \textcolor{red}{LD} & \textcolor{red}{RU} & \textcolor{red}{RU} & \textcolor{red}{LD} & \textcolor{red}{LM} \\
\bottomrule
\end{tabular}
}
\end{table*}

\begin{table*}[ht]
  \caption{Accuracy (top row) and Majority Class Prediction (bottom row) for the \textbf{augmentation and no filtering configuration}. The highest accuracy in each row is \textbf{bolded}. \textcolor{darkgreen}{Green} indicates the predicted majority matches the ground truth, while \textcolor{red}{red} indicates a mismatch.}
\label{tab:results_detailed_2}
\centering
\resizebox{\textwidth}{!}{
\setlength{\tabcolsep}{2pt}
\begin{tabular}{lcccccccccc}
\toprule
Dataset & Conv (Reg) & Conv (Cls) & Conv (Proto) & VGG (Reg) & VGG (Cls) & VGG (Proto) & ResNet (Reg) & ResNet (Cls) & ResNet (Proto) & PCA \\
\midrule
\textbf{Synthetic test} & $96.9 \pm 1.0$ & $82.0 \pm 34.9$ & $12.8 \pm 0.4$ & $12.7 \pm 0.3$ & $12.4 \pm 0.3$ & $96.3 \pm 1.3$ & $98.8 \pm 0.5$ & $\mathbf{99.7} \pm \mathbf{0.1}$ & $14.4 \pm 0.4$ & $36.7 \pm 0.3$ \\[0.2em]
\textbf{1S-LU-6x6-20nm} & $0.0 \pm 0.0$ & $30.3 \pm 15.4$ & $10.1 \pm 7.9$ & $0.0 \pm 0.0$ & $0.0 \pm 0.0$ & $\mathbf{51.8} \pm \mathbf{4.0}$ & $0.0 \pm 0.0$ & $51.0 \pm 9.8$ & $10.4 \pm 6.2$ & $44.4 \pm 0.4$ \\[0.2em]
\textit{Maj. Class} & \textcolor{red}{RU} & \textcolor{darkgreen}{LU} & \textcolor{red}{MD} & \textcolor{red}{RU} & \textcolor{red}{MU} & \textcolor{darkgreen}{LU} & \textcolor{red}{RU} & \textcolor{darkgreen}{LU} & \textcolor{red}{RD} & \textcolor{darkgreen}{LU} \\[0.2em]
\textbf{3S-RM-2x2-20nm} & $87.2 \pm 8.3$ & $48.6 \pm 24.7$ & $15.9 \pm 3.8$ & $24.1 \pm 38.8$ & $20.0 \pm 40.0$ & $69.8 \pm 3.0$ & $\mathbf{88.2} \pm \mathbf{9.0}$ & $81.7 \pm 9.3$ & $14.5 \pm 8.7$ & $75.4 \pm 1.4$ \\[0.2em]
\textit{Maj. Class} & \textcolor{darkgreen}{RM} & \textcolor{darkgreen}{RM} & \textcolor{red}{MD} & \textcolor{red}{RU} & \textcolor{red}{MU} & \textcolor{darkgreen}{RM} & \textcolor{darkgreen}{RM} & \textcolor{darkgreen}{RM} & \textcolor{red}{RD} & \textcolor{darkgreen}{RM} \\[0.2em]
\textbf{5S-RU-6x6-20nm} & $46.4 \pm 9.2$ & $54.1 \pm 28.4$ & $12.5 \pm 3.8$ & $\mathbf{76.0} \pm \mathbf{38.8}$ & $4.0 \pm 8.0$ & $70.7 \pm 1.6$ & $48.0 \pm 4.8$ & $75.0 \pm 2.2$ & $15.0 \pm 8.2$ & $68.1 \pm 1.1$ \\[0.2em]
\textit{Maj. Class} & \textcolor{red}{RM} & \textcolor{darkgreen}{RU} & \textcolor{red}{MD} & \textcolor{darkgreen}{RU} & \textcolor{red}{MU} & \textcolor{darkgreen}{RU} & \textcolor{red}{RM} & \textcolor{darkgreen}{RU} & \textcolor{red}{RD} & \textcolor{darkgreen}{RU} \\[0.2em]
\textbf{6S-LD-4x4-20nm} & $0.0 \pm 0.0$ & $\mathbf{78.1} \pm \mathbf{13.9}$ & $8.8 \pm 7.1$ & $0.0 \pm 0.0$ & $20.0 \pm 40.0$ & $70.8 \pm 5.4$ & $0.0 \pm 0.0$ & $75.9 \pm 2.3$ & $14.8 \pm 6.5$ & $70.2 \pm 1.1$ \\[0.2em]
\textit{Maj. Class} & \textcolor{red}{RM} & \textcolor{darkgreen}{LD} & \textcolor{red}{MD} & \textcolor{red}{RU} & \textcolor{red}{MU} & \textcolor{darkgreen}{LD} & \textcolor{red}{RM} & \textcolor{darkgreen}{LD} & \textcolor{red}{RD} & \textcolor{darkgreen}{LD} \\[0.2em]
\textbf{7S-RD-2x2-20nm} & $0.0 \pm 0.0$ & $56.3 \pm 29.2$ & $6.7 \pm 2.2$ & $0.0 \pm 0.0$ & $20.0 \pm 40.0$ & $75.4 \pm 4.3$ & $0.0 \pm 0.0$ & $\mathbf{84.4} \pm \mathbf{7.6}$ & $19.0 \pm 10.6$ & $67.8 \pm 1.1$ \\[0.2em]
\textit{Maj. Class} & \textcolor{red}{RM} & \textcolor{darkgreen}{RD} & \textcolor{red}{MD} & \textcolor{red}{RU} & \textcolor{red}{MU} & \textcolor{darkgreen}{RD} & \textcolor{red}{RM} & \textcolor{darkgreen}{RD} & \textcolor{darkgreen}{RD} & \textcolor{darkgreen}{RD} \\[0.2em]
\textbf{2S-LU-6x6-50nm} & $0.0 \pm 0.0$ & $9.3 \pm 4.7$ & $11.0 \pm 13.4$ & $0.0 \pm 0.0$ & $0.0 \pm 0.0$ & $\mathbf{15.9} \pm \mathbf{3.4}$ & $0.0 \pm 0.0$ & $9.7 \pm 6.6$ & $11.3 \pm 9.2$ & $12.3 \pm 2.9$ \\[0.2em]
\textit{Maj. Class} & \textcolor{red}{RM} & \textcolor{red}{LD} & \textcolor{red}{RM} & \textcolor{red}{RU} & \textcolor{red}{MU} & \textcolor{darkgreen}{LU} & \textcolor{red}{RM} & \textcolor{red}{RM} & \textcolor{red}{RD} & \textcolor{red}{RM} \\[0.2em]
\textbf{4S-RD-6x6-50nm} & $0.0 \pm 0.0$ & $13.5 \pm 10.6$ & $6.7 \pm 3.4$ & $0.0 \pm 0.0$ & $\mathbf{20.0} \pm \mathbf{40.0}$ & $11.4 \pm 4.0$ & $0.0 \pm 0.0$ & $14.6 \pm 8.5$ & $17.6 \pm 8.4$ & $10.9 \pm 1.5$ \\[0.2em]
\textit{Maj. Class} & \textcolor{red}{RM} & \textcolor{red}{LD} & \textcolor{red}{RM} & \textcolor{red}{RU} & \textcolor{red}{MU} & \textcolor{red}{LU} & \textcolor{red}{RM} & \textcolor{red}{RM} & \textcolor{darkgreen}{RD} & \textcolor{red}{RM} \\[0.2em]
\textbf{1E-RM} & $\mathbf{60.6} \pm \mathbf{18.9}$ & $2.9 \pm 2.3$ & $7.6 \pm 10.8$ & $24.0 \pm 38.8$ & $20.0 \pm 40.0$ & $6.1 \pm 4.0$ & $55.7 \pm 12.4$ & $14.5 \pm 10.9$ & $12.4 \pm 7.7$ & $46.5 \pm 35.3$ \\[0.2em]
\textit{Maj. Class} & \textcolor{darkgreen}{RM} & \textcolor{red}{LD} & \textcolor{red}{MD} & \textcolor{red}{RU} & \textcolor{red}{MU} & \textcolor{red}{LU} & \textcolor{darkgreen}{RM} & \textcolor{red}{LD} & \textcolor{red}{LD} & \textcolor{darkgreen}{RM} \\[0.2em]
\textbf{2E-RM} & $\mathbf{60.6} \pm \mathbf{19.2}$ & $3.0 \pm 2.4$ & $7.3 \pm 10.4$ & $24.0 \pm 38.8$ & $20.0 \pm 40.0$ & $6.1 \pm 3.8$ & $55.9 \pm 12.4$ & $14.5 \pm 11.1$ & $12.5 \pm 7.7$ & $46.5 \pm 35.3$ \\[0.2em]
\textit{Maj. Class} & \textcolor{darkgreen}{RM} & \textcolor{red}{LD} & \textcolor{red}{MD} & \textcolor{red}{RU} & \textcolor{red}{MU} & \textcolor{red}{LU} & \textcolor{darkgreen}{RM} & \textcolor{red}{LD} & \textcolor{red}{LD} & \textcolor{darkgreen}{RM} \\[0.2em]
\textbf{3E-RM} & $\mathbf{59.0} \pm \mathbf{20.9}$ & $3.5 \pm 3.3$ & $6.4 \pm 8.6$ & $24.0 \pm 38.8$ & $20.0 \pm 40.0$ & $6.2 \pm 5.2$ & $57.9 \pm 12.4$ & $13.8 \pm 11.6$ & $11.0 \pm 6.1$ & $47.4 \pm 36.3$ \\[0.2em]
\textit{Maj. Class} & \textcolor{darkgreen}{RM} & \textcolor{red}{LD} & \textcolor{red}{MD} & \textcolor{red}{RU} & \textcolor{red}{MU} & \textcolor{red}{LU} & \textcolor{darkgreen}{RM} & \textcolor{red}{LD} & \textcolor{red}{LD} & \textcolor{darkgreen}{RM} \\[0.2em]
\textbf{4E-RM/RD} & $\mathbf{49.5} \pm \mathbf{20.1}$ & $22.5 \pm 20.1$ & $27.6 \pm 17.8$ & $24.1 \pm 38.8$ & $40.0 \pm 49.0$ & $35.1 \pm 9.0$ & $39.6 \pm 9.3$ & $22.7 \pm 7.8$ & $36.4 \pm 20.9$ & $32.0 \pm 16.4$ \\[0.2em]
\textit{Maj. Class} & \textcolor{red}{RU} & \textcolor{red}{LD} & \textcolor{red}{RU} & \textcolor{red}{RU} & \textcolor{red}{MU} & \textcolor{darkgreen}{RD} & \textcolor{red}{RU} & \textcolor{red}{LU} & \textcolor{darkgreen}{RM} & \textcolor{darkgreen}{RM} \\[0.2em]
\textbf{5E-RM/RD} & $46.8 \pm 23.5$ & $12.1 \pm 13.4$ & $\mathbf{49.2} \pm \mathbf{27.0}$ & $24.1 \pm 38.8$ & $40.0 \pm 49.0$ & $34.3 \pm 7.9$ & $27.4 \pm 9.2$ & $18.3 \pm 7.1$ & $27.7 \pm 12.7$ & $43.7 \pm 31.0$ \\[0.2em]
\textit{Maj. Class} & \textcolor{red}{RU} & \textcolor{red}{LD} & \textcolor{darkgreen}{RM} & \textcolor{red}{RU} & \textcolor{red}{MU} & \textcolor{darkgreen}{RD} & \textcolor{red}{RU} & \textcolor{red}{LD} & \textcolor{red}{RU} & \textcolor{darkgreen}{RM} \\[0.2em]
\bottomrule
\end{tabular}
}
\end{table*}

\begin{table*}[ht]
  \caption{Accuracy (top row) and Majority Class Prediction (bottom row) for the \textbf{augmentation and filtering configuration}. The highest accuracy in each row is \textbf{bolded}. \textcolor{darkgreen}{Green} indicates the predicted majority matches the ground truth, while \textcolor{red}{red} indicates a mismatch.}
\label{tab:results_detailed_3}
\centering
\resizebox{\textwidth}{!}{
\setlength{\tabcolsep}{2pt}
\begin{tabular}{lcccccccccc}
\toprule
Dataset & Conv (Reg) & Conv (Cls) & Conv (Proto) & VGG (Reg) & VGG (Cls) & VGG (Proto) & ResNet (Reg) & ResNet (Cls) & ResNet (Proto) & PCA \\
\midrule
\textbf{Synthetic test} & $96.0 \pm 0.9$ & $99.3 \pm 0.1$ & $79.9 \pm 6.1$ & $98.7 \pm 0.5$ & $45.0 \pm 40.3$ & $94.5 \pm 4.3$ & $98.4 \pm 0.6$ & $\mathbf{99.5} \pm \mathbf{0.3}$ & $92.4 \pm 2.1$ & $38.8 \pm 0.4$ \\[0.2em]
\textbf{1S-LU-6x6-20nm} & $0.0 \pm 0.0$ & $26.5 \pm 9.0$ & $15.2 \pm 2.5$ & $0.0 \pm 0.0$ & $6.6 \pm 8.5$ & $\mathbf{32.4} \pm \mathbf{5.8}$ & $0.0 \pm 0.0$ & $24.9 \pm 2.5$ & $28.2 \pm 3.0$ & $25.6 \pm 0.7$ \\[0.2em]
\textit{Maj. Class} & \textcolor{red}{RU} & \textcolor{darkgreen}{LU} & \textcolor{darkgreen}{LU} & \textcolor{red}{RU} & \textcolor{red}{MD} & \textcolor{darkgreen}{LU} & \textcolor{red}{RU} & \textcolor{darkgreen}{LU} & \textcolor{darkgreen}{LU} & \textcolor{darkgreen}{LU} \\[0.2em]
\textbf{3S-RM-2x2-20nm} & $62.7 \pm 9.5$ & $32.0 \pm 2.9$ & $26.3 \pm 3.0$ & $\mathbf{73.8} \pm \mathbf{8.8}$ & $18.4 \pm 22.4$ & $35.7 \pm 7.4$ & $65.6 \pm 7.6$ & $42.8 \pm 3.3$ & $34.0 \pm 2.9$ & $36.6 \pm 0.5$ \\[0.2em]
\textit{Maj. Class} & \textcolor{darkgreen}{RM} & \textcolor{darkgreen}{RM} & \textcolor{darkgreen}{RM} & \textcolor{darkgreen}{RM} & \textcolor{red}{MD} & \textcolor{darkgreen}{RM} & \textcolor{darkgreen}{RM} & \textcolor{darkgreen}{RM} & \textcolor{darkgreen}{RM} & \textcolor{darkgreen}{RM} \\[0.2em]
\textbf{5S-RU-6x6-20nm} & $50.8 \pm 9.8$ & $33.1 \pm 1.5$ & $23.5 \pm 4.5$ & $48.8 \pm 13.8$ & $14.8 \pm 18.4$ & $36.6 \pm 8.2$ & $\mathbf{53.2} \pm \mathbf{8.7}$ & $40.4 \pm 1.8$ & $33.4 \pm 2.2$ & $36.3 \pm 1.1$ \\[0.2em]
\textit{Maj. Class} & \textcolor{darkgreen}{RU} & \textcolor{darkgreen}{RU} & \textcolor{darkgreen}{RU} & \textcolor{red}{RM} & \textcolor{red}{MD} & \textcolor{darkgreen}{RU} & \textcolor{darkgreen}{RU} & \textcolor{darkgreen}{RU} & \textcolor{darkgreen}{RU} & \textcolor{darkgreen}{RU} \\[0.2em]
\textbf{6S-LD-4x4-20nm} & $0.0 \pm 0.0$ & $39.9 \pm 8.8$ & $22.4 \pm 4.0$ & $0.0 \pm 0.0$ & $15.4 \pm 13.8$ & $42.0 \pm 5.4$ & $0.0 \pm 0.0$ & $\mathbf{42.9} \pm \mathbf{5.3}$ & $35.5 \pm 2.2$ & $34.5 \pm 0.8$ \\[0.2em]
\textit{Maj. Class} & \textcolor{red}{RU} & \textcolor{darkgreen}{LD} & \textcolor{darkgreen}{LD} & \textcolor{red}{RM} & \textcolor{red}{MD} & \textcolor{darkgreen}{LD} & \textcolor{red}{RU} & \textcolor{darkgreen}{LD} & \textcolor{darkgreen}{LD} & \textcolor{darkgreen}{LD} \\[0.2em]
\textbf{7S-RD-2x2-20nm} & $0.0 \pm 0.0$ & $33.6 \pm 3.3$ & $24.8 \pm 3.3$ & $0.0 \pm 0.0$ & $19.8 \pm 24.3$ & $\mathbf{42.0} \pm \mathbf{7.6}$ & $0.0 \pm 0.0$ & $41.7 \pm 1.9$ & $37.7 \pm 5.7$ & $32.4 \pm 0.5$ \\[0.2em]
\textit{Maj. Class} & \textcolor{red}{RM} & \textcolor{darkgreen}{RD} & \textcolor{darkgreen}{RD} & \textcolor{red}{RM} & \textcolor{red}{MD} & \textcolor{darkgreen}{RD} & \textcolor{red}{RM} & \textcolor{darkgreen}{RD} & \textcolor{darkgreen}{RD} & \textcolor{darkgreen}{RD} \\[0.2em]
\textbf{2S-LU-6x6-50nm} & $0.0 \pm 0.0$ & $\mathbf{18.1} \pm \mathbf{9.6}$ & $6.8 \pm 3.4$ & $0.0 \pm 0.0$ & $1.3 \pm 1.7$ & $13.2 \pm 4.6$ & $0.0 \pm 0.0$ & $5.9 \pm 1.9$ & $14.5 \pm 4.8$ & $16.0 \pm 3.4$ \\[0.2em]
\textit{Maj. Class} & \textcolor{red}{RM} & \textcolor{red}{LD} & \textcolor{red}{RM} & \textcolor{red}{RM} & \textcolor{red}{MD} & \textcolor{red}{MD} & \textcolor{red}{RU} & \textcolor{red}{MD} & \textcolor{red}{MD} & \textcolor{red}{RM} \\[0.2em]
\textbf{4S-RD-6x6-50nm} & $0.0 \pm 0.0$ & $\mathbf{15.0} \pm \mathbf{3.4}$ & $10.2 \pm 2.0$ & $0.0 \pm 0.0$ & $8.3 \pm 12.4$ & $10.5 \pm 2.7$ & $0.0 \pm 0.0$ & $10.3 \pm 2.5$ & $13.2 \pm 6.6$ & $11.5 \pm 2.0$ \\[0.2em]
\textit{Maj. Class} & \textcolor{red}{RM} & \textcolor{red}{LU} & \textcolor{red}{RM} & \textcolor{red}{RM} & \textcolor{red}{MD} & \textcolor{red}{MU} & \textcolor{red}{RU} & \textcolor{red}{MD} & \textcolor{red}{MD} & \textcolor{red}{RM} \\[0.2em]
\textbf{1E-RM} & $53.6 \pm 25.6$ & $8.5 \pm 6.6$ & $39.8 \pm 9.3$ & $\mathbf{53.8} \pm \mathbf{12.3}$ & $18.2 \pm 30.5$ & $5.6 \pm 5.3$ & $40.9 \pm 18.6$ & $9.7 \pm 5.0$ & $3.6 \pm 2.6$ & $35.5 \pm 32.4$ \\[0.2em]
\textit{Maj. Class} & \textcolor{darkgreen}{RM} & \textcolor{red}{LM} & \textcolor{darkgreen}{RM} & \textcolor{darkgreen}{RM} & \textcolor{red}{MD} & \textcolor{red}{RD} & \textcolor{red}{RU} & \textcolor{red}{MD} & \textcolor{red}{LD} & \textcolor{darkgreen}{RM} \\[0.2em]
\textbf{2E-RM} & $53.6 \pm 25.8$ & $8.5 \pm 6.6$ & $39.9 \pm 9.7$ & $\mathbf{53.7} \pm \mathbf{12.2}$ & $18.2 \pm 30.3$ & $5.5 \pm 5.2$ & $40.6 \pm 18.6$ & $9.9 \pm 5.1$ & $3.7 \pm 2.5$ & $35.5 \pm 32.4$ \\[0.2em]
\textit{Maj. Class} & \textcolor{darkgreen}{RM} & \textcolor{red}{LM} & \textcolor{darkgreen}{RM} & \textcolor{darkgreen}{RM} & \textcolor{red}{MD} & \textcolor{red}{RD} & \textcolor{red}{RU} & \textcolor{red}{MD} & \textcolor{red}{LD} & \textcolor{darkgreen}{RM} \\[0.2em]
\textbf{3E-RM} & $51.3 \pm 27.0$ & $8.6 \pm 5.1$ & $40.3 \pm 9.3$ & $\mathbf{54.6} \pm \mathbf{9.1}$ & $18.2 \pm 30.3$ & $4.8 \pm 4.5$ & $41.9 \pm 15.4$ & $9.5 \pm 5.3$ & $3.4 \pm 2.3$ & $38.6 \pm 33.4$ \\[0.2em]
\textit{Maj. Class} & \textcolor{darkgreen}{RM} & \textcolor{red}{LM} & \textcolor{darkgreen}{RM} & \textcolor{darkgreen}{RM} & \textcolor{red}{MD} & \textcolor{red}{MU} & \textcolor{red}{RU} & \textcolor{red}{MD} & \textcolor{red}{LD} & \textcolor{darkgreen}{RM} \\[0.2em]
\textbf{4E-RM/RD} & $53.7 \pm 22.0$ & $2.7 \pm 1.8$ & $\mathbf{71.4} \pm \mathbf{15.6}$ & $54.6 \pm 11.3$ & $24.0 \pm 22.0$ & $21.5 \pm 14.9$ & $22.4 \pm 17.8$ & $6.8 \pm 5.1$ & $17.1 \pm 12.3$ & $37.2 \pm 20.1$ \\[0.2em]
\textit{Maj. Class} & \textcolor{darkgreen}{RM} & \textcolor{red}{MD} & \textcolor{darkgreen}{RM} & \textcolor{darkgreen}{RM} & \textcolor{red}{MD} & \textcolor{red}{MU} & \textcolor{red}{RU} & \textcolor{red}{LM} & \textcolor{red}{LU} & \textcolor{darkgreen}{RM} \\[0.2em]
\textbf{5E-RM/RD} & $26.9 \pm 20.2$ & $3.9 \pm 4.4$ & $\mathbf{58.6} \pm \mathbf{18.6}$ & $42.1 \pm 7.9$ & $21.2 \pm 19.6$ & $18.5 \pm 12.9$ & $18.1 \pm 13.8$ & $12.3 \pm 8.6$ & $20.7 \pm 12.6$ & $46.6 \pm 21.1$ \\[0.2em]
\textit{Maj. Class} & \textcolor{red}{RU} & \textcolor{red}{MD} & \textcolor{darkgreen}{RM} & \textcolor{red}{RU} & \textcolor{red}{MD} & \textcolor{red}{MU} & \textcolor{red}{RU} & \textcolor{red}{MD} & \textcolor{red}{LU} & \textcolor{darkgreen}{RM} \\[0.2em]
\bottomrule
\end{tabular}
}
\end{table*}

The models perform well on the synthetic test set (see first row in each Table), when no augmentation and no filtering are applied, with many models achieving accuracies above 99\%. This indicates that all approaches are highly effective at classifying data that is structurally identical to the training data, since the synthetic test set is made of 10\% of diffraction patterns from the same structures on which the models are trained (i.e., 10\% of $D_t$).

A different pattern emerges when testing against other synthetic datasets. All models perform worse on at least some of the other synthetic data, but most models still manage to predict the correct majority class on synthetic datasets with the same 20~nm thickness. A key finding is that all methods fail for the thicker 50 nm models. This is attributed to a physical limitation: the increased complexity of dynamical electron scattering fundamentally alters the image contrast. The signal transitions from a simple projection of the structure into a non-linear, integrated one, breaking the core assumption of the thin-sample simulations on which the models were trained. In contrast, models tend to perform well on larger lateral models ($4\times4$ or $6\times6$), demonstrating their efficacy for larger-scale systems. This is particularly relevant because experimental images are rarely obtained from a very small $2\times2$ unit cell area.

The performance on the experimental datasets is poor when \textit{no augmentation and no filtering} are applied. This result points to a substantial domain gap between the idealized synthetic training data and the noisy, real-world experimental data. The best results are achieved by VGG (Reg), achieving accuracies of more than 40\% across all experimental sets, yet still failing to predict the correct majority class. In fact, the results of all three regression approaches are most likely invalid since, excluding the synthetic test set, these approaches always predict just two out of 8 possible classes for all diffraction patterns in all test sets, RU and RM. This strong bias toward two classes and zero variance suggests the regression loss is failing to learn the underlying multi-class polarization distribution, effectively becoming a trivial predictor for most of these datasets.

In the \textit{augmentation and no filtering} configuration, a failure to converge is observed in several models, specifically Conv (Proto), VGG (Reg), VGG (Cls), and ResNet (Proto). The non-convergence is indicated by a drastic reduction in performance on the synthetic test set, where these models achieve an accuracy of approximately 12.5\%, a figure consistent with a trivial majority classifier. Interestingly, for the subset of classifiers that successfully converge, there is a visible improvement in performance on experimental datasets. This suggests that while augmentation may introduce training complexities, it plays a vital role in bridging the domain gap between synthetic and experimental data. Notably, PCA correctly identifies the majority classes for all 20~nm thick synthetic datasets and all experimental datasets. However, the accompanying large standard deviation introduces doubts regarding the robustness of this approach.

The introduction of train set filtering, specifically targeting examples with above-average polarization magnitude, restores convergence across all models when used in tandem with augmentation. In the \textit{augmentation and filtering}  configuration, Conv (Proto) and PCA demonstrate superior performance by correctly predicting all majority classes for experimental datasets and four out of six for synthetic datasets. This success indicates that filtering assists convergence while retaining enough diversity to mitigate the domain gap. However, these improvements are tempered by significant drawbacks. The train set preprocessing markedly decreases accuracy on synthetic datasets for both leading models. Furthermore, performance remains relatively low, at approximately 40\%, for the three experimental datasets involving a single (RM) solution (1E-RM, 2E-RM, and 3E-RM). Most critically, the consistently large standard deviations suggest that the results remain unreliable. The high degree of variance indicates that the predicted majority classes and overall model performance are influenced by the specific random seed used, rendering the consistency of these correct predictions statistically questionable.

Another interesting observation is that the usage of models pretrained on ImageNet (VGG and ResNet) does not help with the performance. These models either produce trivial predictions (i.e., they always predict the same class) when trained with the regression objective, or produce wrong majority class predictions, when trained with the classification or prototype representation objective, suggesting that the fine-tuned models fail to learn any meaningful features for inference on the experimental sets.

\subsection{Classification Error Analysis}

\begin{figure}[htb]
  \centering
  \includegraphics[width=1.0\textwidth]{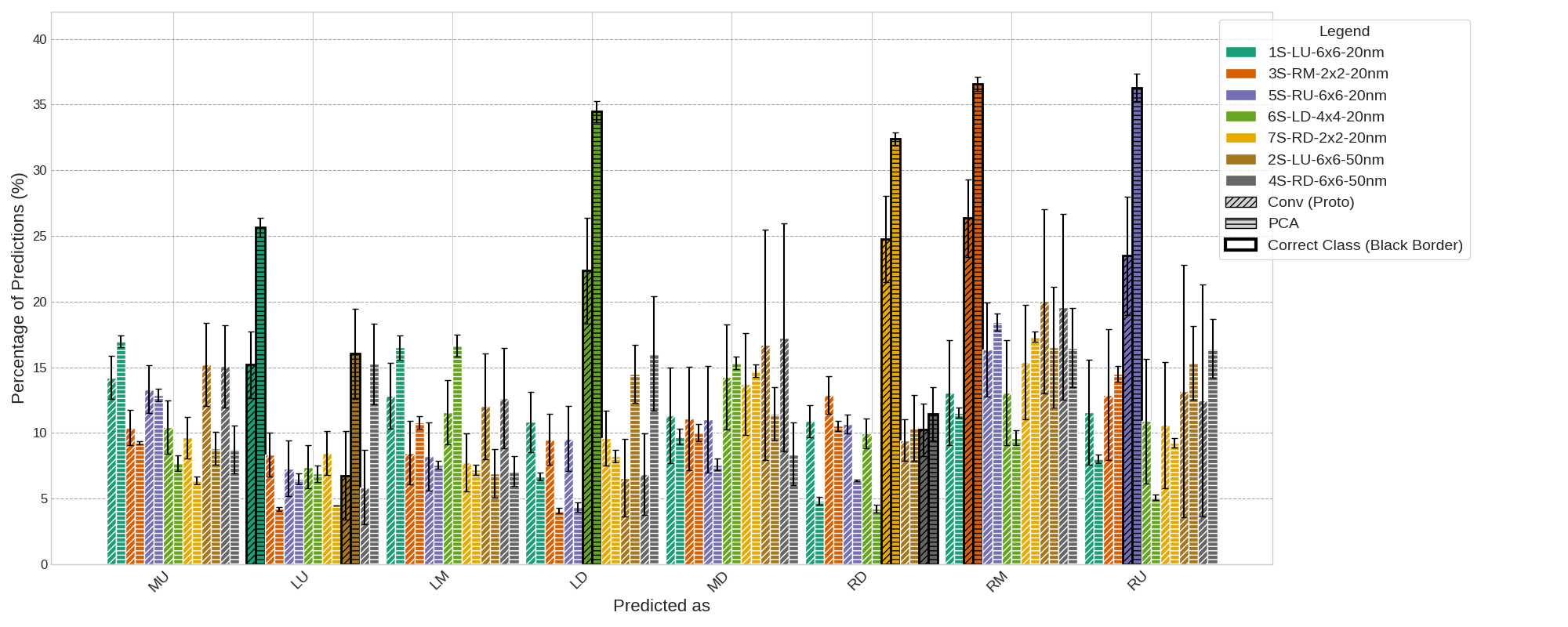}
  \caption{Prediction distribution on the synthetic test datasets for the Conv (Proto) and PCA methods with standard deviation across 5 seeds.}
  \label{fig:bar-plot-synth}
\end{figure}

In our error analysis, we closely examine the performance of the two best methods, Conv (Proto) and PCA  in the \textit{augmentation and filtering} configuration. Figure \ref{fig:bar-plot-synth} shows prediction distribution on the synthetic datasets for the Conv (Proto) and PCA methods with standard deviation across 5 seeds. Note that the prediction distributions of the two methods are somewhat correlated across synthetic test sets in a sense that the classes neighboring the correct classes tend to get a larger share of missclassifications than other classes, when the majority class is correctly predicted (i.e., on all datasets but the 2S-LU-6x6-50nm and 4S-RD-6x6-50nm). This suggests that both models to some extent consider the fact that the physical polarization directions are continuous and spatially related. While this is not surprising for the Conv (Proto) approach, where the training loss explicitly considers geometric relationships between different classes, it is interesting that a similar pattern is also observed in the PCA approach, where different classes are modeled as independent. When it comes to two datasets with a missclassified majority class (2S-LU-6x6-50nm and 4S-RD-6x6-50nm), no specific general pattern is observed. 

\begin{figure}[htb]
  \centering
  \includegraphics[width=1.0\textwidth]{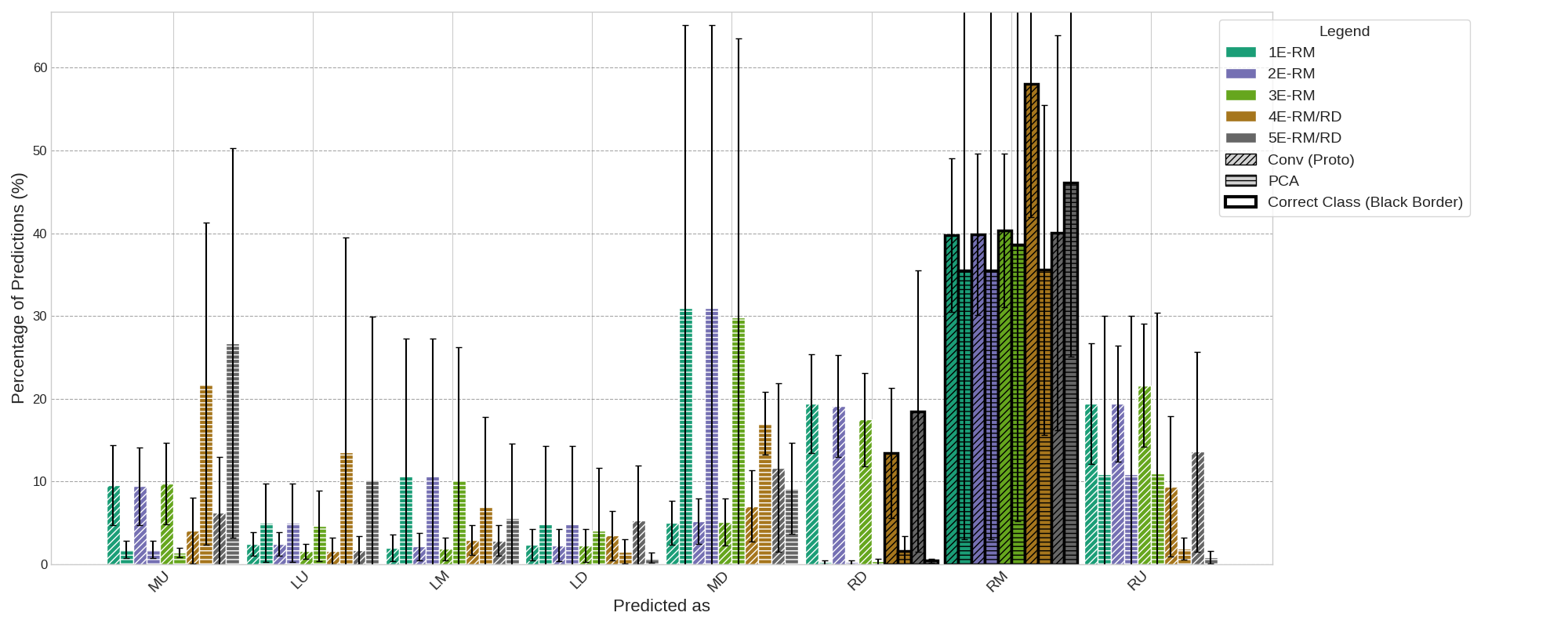}
  \caption{Prediction distribution on the experimental test datasets for the Conv (Proto) and PCA methods with standard deviation across 5 seeds.}
  \label{fig:bar-plot-exp}
\end{figure}

Figure \ref{fig:bar-plot-exp} shows prediction distribution on the experimental test datasets for the Conv (Proto) and PCA methods with standard deviation across 5 seeds. Here, we observe a weaker correlation between the PCA and the Conv (Proto) models when it comes to the distribution of missclassifications. For example, for the 4E-RM/RD dataset, PCA missclassifies about 21\% of diffraction patterns into the MU class, while Conv (Proto) classifies only around 8\% of examples as MU. Also, the classes neighboring the correct RM class (the RD and RU classes) do not always contain the majority of missclassifications.

\begin{figure}[htb]
  \centering
  \includegraphics[width=1.0\textwidth]{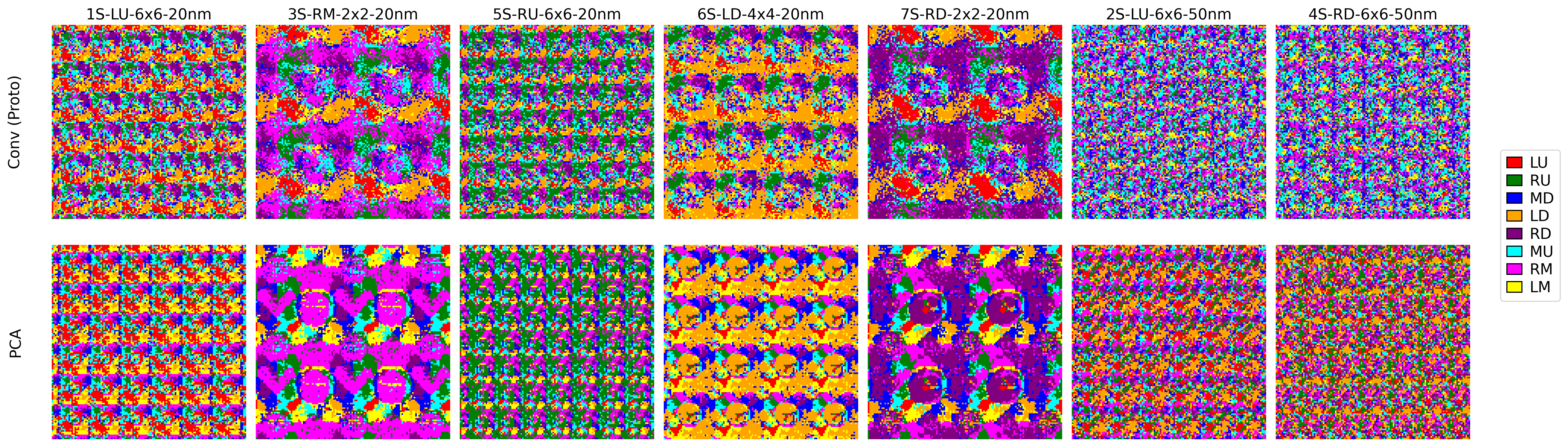}
  \caption{Polarization classification (different color per each class) for each diffraction pattern in a $128\times128$ structure across all synthetic structures. We take a majority vote across five seeds for each prediction.}
  \label{fig:heat-map-synth}
\end{figure}

\begin{figure}[htb]
  \centering
  \includegraphics[width=1.0\textwidth]{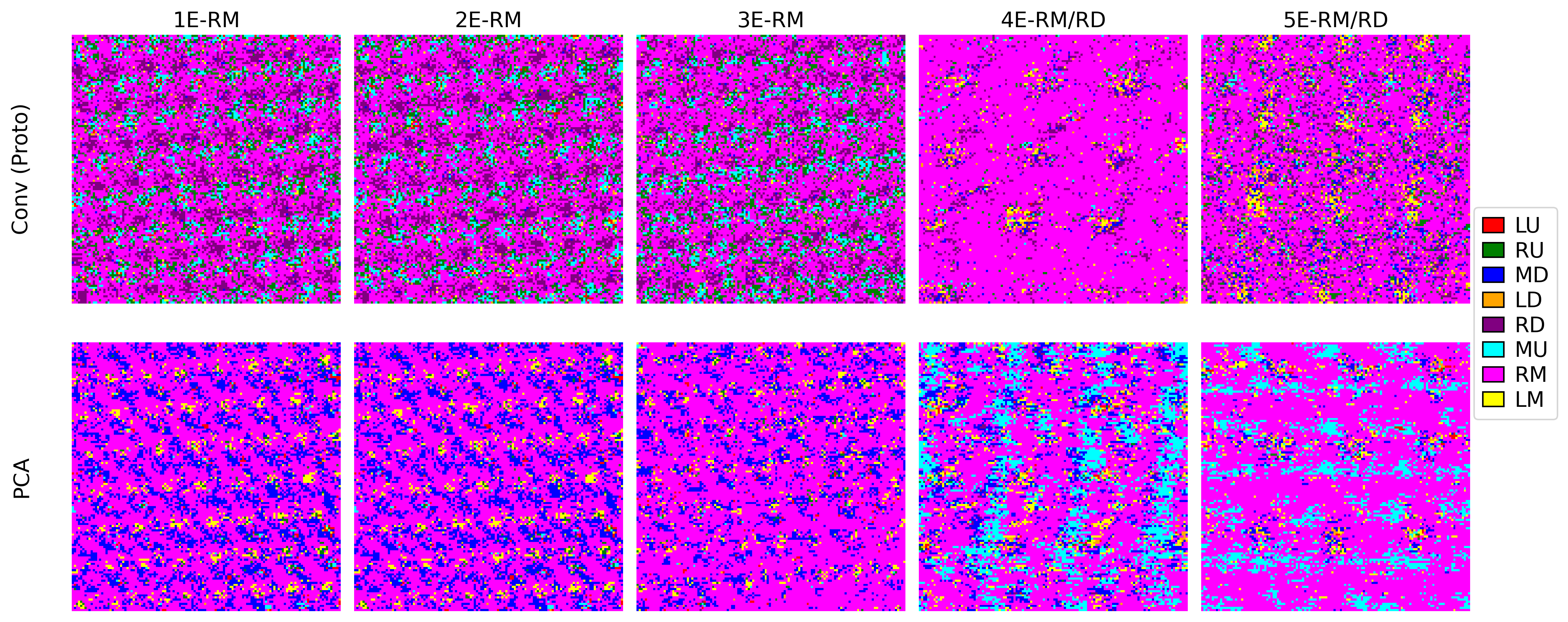}
  \caption{Polarization classification (different color per each class) for each diffraction pattern in a $128\times128$ structure across all experimental structures for PCA and Conv (Proto). We take a majority vote across five seeds for each prediction.}
  \label{fig:heat-map-exp}
\end{figure}

\begin{figure}[htb]
  \centering
  \includegraphics[width=1.0\textwidth]{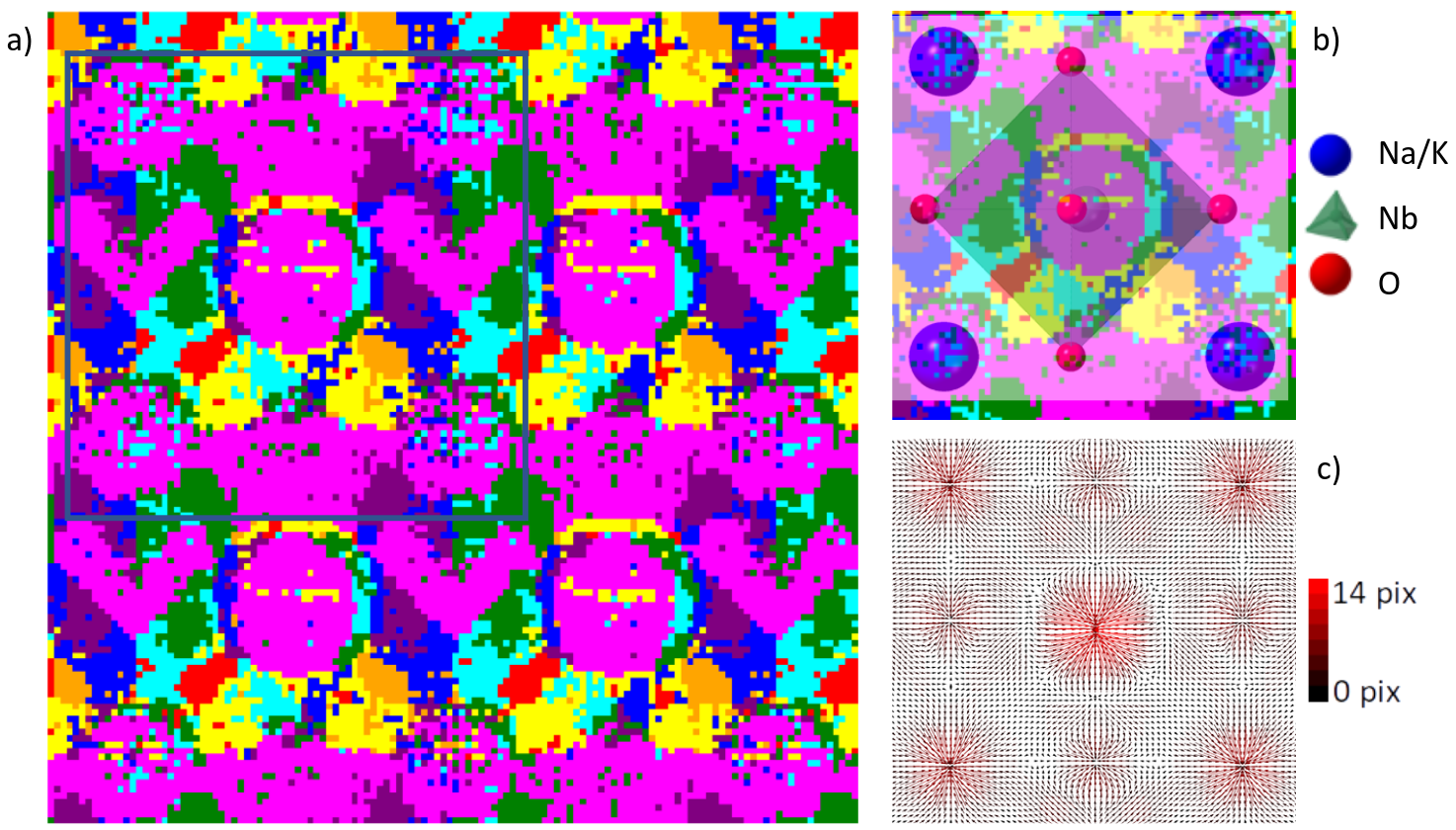}
  \caption{a) Polarization classification map for the 3S-RM-2x2-20nm. b) Unit cell area, with an overlaid KNN structural model and c) the corresponding electron beam shift vector field calculated from the center of mass (CoM).}
  \label{fig:heat-map-overlap}
\end{figure}

We also analyze the specific diffraction pattern predictions within the $128\times128$ grid structure to identify misclassification patterns that may shed light on model behavior and 4D-STEM KNN structure representations. We provide a per-pixel determination of the in-plane polarization direction in Figures \ref{fig:heat-map-synth} and \ref{fig:heat-map-exp}. Figure \ref{fig:heat-map-synth} shows polarization classification (different color per each class) for each diffraction pattern in a $128\times128$ structure across all synthetic datasets for PCA and Conv (Proto). We take a majority vote across five seeds for each diffraction prediction and each class is represented by a different color. 

For the 3S and 7S samples, a substantial fraction of pixels correctly resolves the expected RM or RD configuration, respectively. However, Figure \ref{fig:heat-map-synth} shows the solutions become unreliable in the inter-columnar regions. This indicates that the diffraction signals from these areas lack meaningful polarization information, yielding random assignments. These unreliable regions correlate strongly with areas exhibiting a low-magnitude electron beam shift vector, as derived from the CoM analysis. While the same correlation is also observed for the larger lateral periodicity models (1S, 5S, and 6S), this trend does not hold for the thicker (50 nm) samples (2S and 4S), where the majority class is not correct. This suggests that not all diffraction patterns carry equally relevant information about the polarization direction. The most critical information is mostly contained in the diffractions in the vicinity of the atom columns, which correspond to areas exhibiting the largest CoM shifts (see Figure \ref{fig:heat-map-overlap}), upon which the algorithm bases its final overall solution (through majority voting).

Figure \ref{fig:heat-map-exp} shows polarization classification (different color per each class) for each diffraction pattern in a $128\times128$ structure across all experimental structures for PCA and Conv (Proto). The polarization mapping results across the five experimental datasets (1E–5E) demonstrate the model’s adaptability to varying spatial scales. While datasets 1E, 2E, and 3E encompass a larger number of unit cells compared to the more localized field-of-views in 4E and 5E, the lateral size of the scan area did not impede the convergence or accuracy of the polarization detection. This consistency across different sampling densities aligns with our preliminary simulations, which suggested that the solution space remains stable regardless of the number of probed unit cells. Furthermore, although observable scan drift was present during acquisition, its impact on the final polarization vectors appears negligible, likely due to the localized nature of the diffraction pattern analysis. For 4E and 5E samples, where the inversion symmetry breaking allows for multiple potential polarization states, the algorithm most likely converges on the solution with the higher magnitude.

\subsection{Anomaly Detection}

We additionally conduct a preliminary feasibility study where we investigate whether the models trained for the task of spontaneous polarization detection can be used to identify deviations in crystal structure. Because labeled experimental defect data are not available, we use a synthetic dataset in which structural defects and noise are artificially introduced. We hypothesize that the magnitude and direction of polarization may change near structural defects for several reasons:  

\begin{itemize}
\item Defects such as dislocations, vacancies, or dopants introduce local strain. In ferroelectrics like KNN, strain is strongly coupled to polarization (via piezoelectric, electrostrictive, and flexoelectric effects), so lattice distortions can enhance local dipole moments.
\item Charged defects (such as oxygen vacancies or aliovalent dopants) can locally disrupt charge balance, potentially leading to stronger local electric fields or reorganization of local dipoles, and thus altering the net polarization magnitude nearby.
\item Structural defects often break local symmetry, allowing polarization components that are otherwise suppressed in the pristine structure, which could result in enhanced polarization magnitudes or the emergence of new polarization directions.
\end{itemize}

As explained above, the classifiers are trained to produce a polarization label for each diffraction pattern in a $128 \times 128$ grid structure. If structural defects affect polarization in specific regions of the KNN structure, the assumption is that classifiers will ``missclassify'' diffraction patterns covering those regions. Due to availability of positional information for each diffraction pattern (i.e., we know the position of a specific diffraction pattern in a $128 \times 128$ grid), we can identify regions inside the grid with a different pattern of missclassification. These regions should, according to our hypothesis, have a large intersection with the defected regions.

\begin{figure}[htb]
  \centering
  \includegraphics[width=1.0\textwidth]{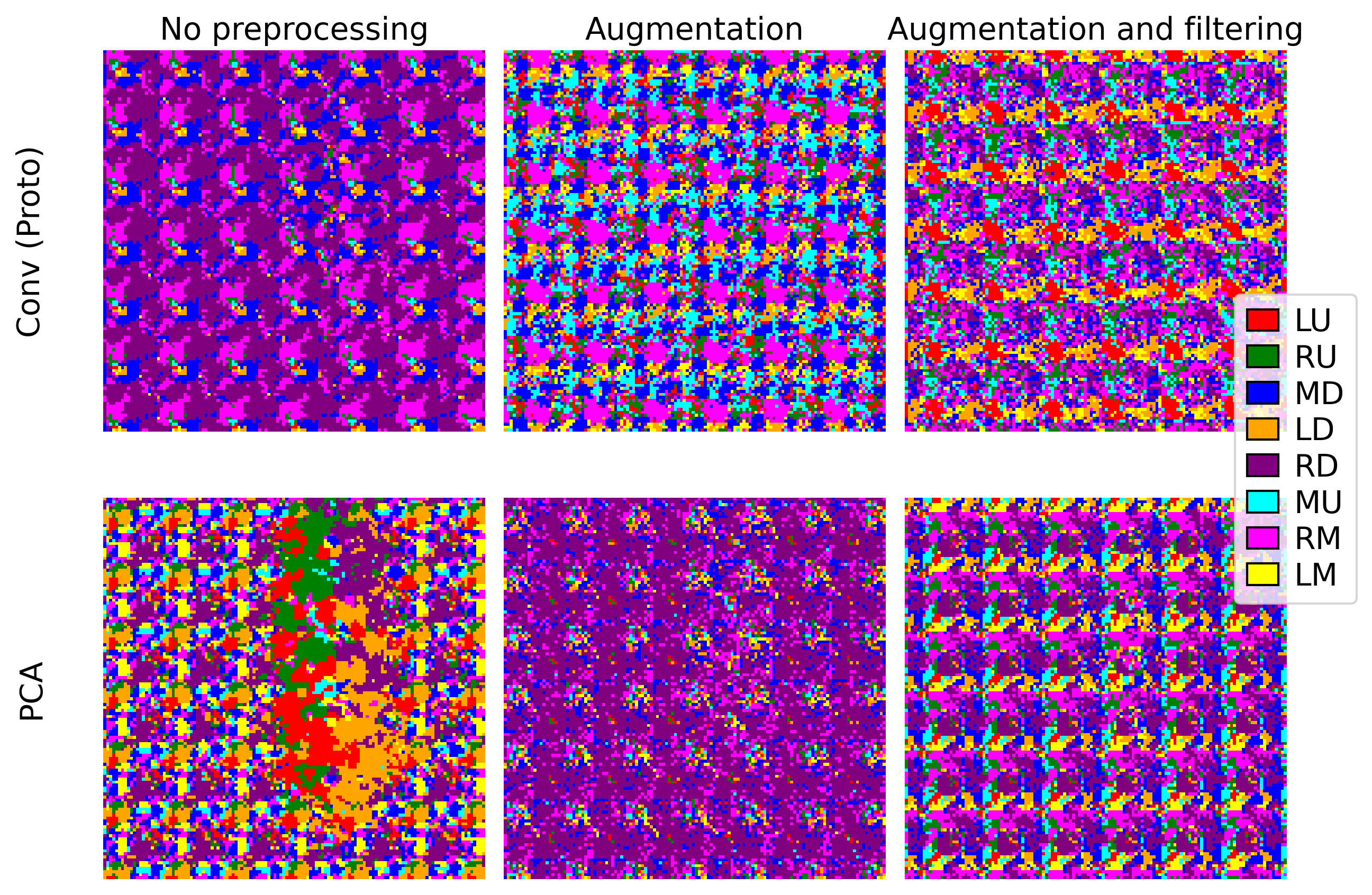}
  \caption{Polarization classification (different color per each class) for each diffraction pattern in a $128 \times 128$ RD structure with defect across three configurations for Conv (Proto) and PCA approaches.}
  \label{with+without-anomalies}
\end{figure}

Figure \ref{with+without-anomalies} shows polarization classification results (different colors for each diffraction pattern in a $128 \times 128$ structure) with a simulated defect across three configurations (left -- no augmentation and no filtering, middle -- augmentation and no filtering, and right -- augmentation and filtering) for the PCA and Conv (Proto) approaches. With the Conv (Proto) approach, the anomaly is not visible in any of the three configurations. On the other hand, PCA managed to successfully identify the defect site, because it causes detectable changes in polarization, as evidenced by DFT calculations. Although the defect was introduced in a single atomic column, the resulting change in polarization extends several unit cells away.
In particular, DFT calculations predict that an oxygen vacancy induces a pronounced relaxation pattern involving at least three neighboring Nb ions along the linear chain containing the defect. In the pristine structure, this chain follows the sequence Nb--O$\cdots$Nb--O$\cdots$Nb--O, where Nb--O separations alternate between shorter (--) and longer ($\cdots$) distances. Around the vacancy, the neighboring Nb ions relax away from the vacancy along the chain axis. The closest Nb ion exhibits the largest displacement of approximately 0.4~\AA. The second Nb ion along the same chain direction relaxes by about 0.3~\AA, while the Nb ion on the opposite side of the vacancy relaxes by slightly more than 0.1~\AA.
Note that a clearly visible distortion in the misclassification pattern appears when no preprocessing is applied. With applied augmentation, the distortion becomes much less visible and disappears when both augmentation and filtering are applied. This suggests that preprocessing, while effective for reducing the domain gap between synthetic training and experimental test structures, reduces the precision of defect detection. Consequently, the feasibility of defect detection on experimental datasets --- where preprocessing is even more critical for accurate polarization classification --- remains uncertain. Furthermore, the divergence in detection capability suggests that PCA features remain more closely aligned with traditional CoM features, whereas Conv (Proto) relies on learned features that are largely uncorrelated with the specific signals produced by the anomaly.

\section{Discussion}

In this study, we have explored different supervised machine learning approaches for determination of spontaneous polarization in perovskite ferroelectrics utilizing 4D-STEM datasets. For polarization direction mapping, we tested three neural architectures, ResNet, VGG, and a custom convolutional neural network (CNN), and a more traditional approach using k-Nearest Neighbors (k-NN) classifier with Principal Component Analysis (PCA). We also explore three distinct training paradigms, standard classification, regression and a prototype representation approach. While most models perform well on synthetic datasets with the 20~nm sample thickness, we find that a tailored prototype-based approach is the only neural approach that can partially reduce the domain gap between the synthetic training datasets and experimental test datasets, if extensive augmentation and filtering is applied on the train set. 

Overall, the PCA-based approach showed the most robust performance, predicting correct polarization majority classes for all experimental datasets in two out of three preprocessing configurations. However, the low mean accuracy and high variance of the best two approaches (PCA and Conv (Proto)) on the experimental datasets indicate that these methods are not yet reliable for fine-grained, pixel-level classification. In future work, we plan to test neural models pretrained on microscopic images to help reduce the domain gap, and further evaluate the generalizability of the two best approaches on additional experimental data.

A fundamental limitation of this study is that the analysis was designed for thinner models; as a result, it fails for thicker models and experimental specimens (above 20 nm thickness), due to multiple scattering of electrons obscuring displacement signals. This limitation is particularly important for ferroelectric materials, where understanding both the net polarization direction and the precise position and atomic configuration of defects (e.g., point defects, vacancy clusters) is essential for governing functional properties. 

\section{Data Availability}

The data that support the findings of this study are available at \url{https://kt-cloud.ijs.si/index.php/s/yKqWe26tAF5tgC5}.

\section{Code Availability}

The code for polarization classification, along with detailed instructions, is available at \url{https://github.com/matejMartinc/4D-STEM-polarization-mapping}.

%
%
%
\bibliographystyle{naturemag}
\bibliography{bibliography}

@article{10.1017/S1431927619000497,
    author = {Ophus, Colin},
    title = {Four-Dimensional Scanning Transmission Electron Microscopy (4D-STEM): From Scanning Nanodiffraction to Ptychography and Beyond},
    journal = {Microscopy and Microanalysis},
    volume = {25},
    number = {3},
    pages = {563-582},
    year = {2019},
    month = {06},
    issn = {1431-9276},
    doi = {10.1017/S1431927619000497},
    url = {https://doi.org/10.1017/S1431927619000497},
    eprint = {https://academic.oup.com/mam/article-pdf/25/3/563/47996685/mam0563.pdf},
}

@inproceedings{sadri2025unsupervised,
  title={Unsupervised Deep Denoising for Low-Dose 4D-STEM},
  author={Sadri, Alireza and Petersen, Timothy and Terzoudis-Lumsden, Emmanuel WC and Esser, Bryan David and Etheridge, Joanne and Findlay, Scott D},
  booktitle={13th Asia Pacific Microscopy Congress 2025 (APMC13)},
  pages={205},
  year={2025},
  organization={ScienceOpen}
}

@article{zhu2024structural,
  title={Structural degeneracy and formation of crystallographic domains in epitaxial LaFeO3 films revealed by machine-learning assisted 4D-STEM},
  author={Zhu, Menglin and Lanier, Joseph and Flores, Jose and da Cruz Pinha Barbosa, Victor and Russell, Daniel and Haight, Becky and Woodward, Patrick M and Yang, Fengyuan and Hwang, Jinwoo},
  journal={Scientific Reports},
  volume={14},
  number={1},
  pages={4198},
  year={2024},
  publisher={Nature Publishing Group UK London}
}

@article{10.1093/micmic/ozad067.1015,
    author = {Yoo, Timothy and Hershkovitz, Eitan and Pu, Xiaofei and He, Lingfeng and Kim, Honggyu},
    title = {Conjoining Simple Supervised and Unsupervised Machine Learning Methods with 4D-STEM to Identify Complex Nanostructures},
    journal = {Microscopy and Microanalysis},
    volume = {29},
    number = {Supplement\_1},
    pages = {1959-1960},
    year = {2023},
    month = {07},
    issn = {1431-9276},
    doi = {10.1093/micmic/ozad067.1015},
    url = {https://doi.org/10.1093/micmic/ozad067.1015},
}

@article{roccapriore2022automated,
  title={Automated experiment in 4D-STEM: exploring emergent physics and structural behaviors},
  author={Roccapriore, Kevin M and Dyck, Ondrej and Oxley, Mark P and Ziatdinov, Maxim and Kalinin, Sergei V},
  journal={ACS nano},
  volume={16},
  number={5},
  pages={7605--7614},
  year={2022},
  publisher={ACS Publications}
}

@book{cao2021new,
  title={New Imaging Methods with 4D-STEM: Quantitative Mapping of Fields, Polarity, Tilt, and Phase},
  author={Cao, Michael Chen},
  year={2021},
  publisher={Cornell University}
}

@article{kimoto2024unsupervised,
  title={Unsupervised machine learning combined with 4D scanning transmission electron microscopy for bimodal nanostructural analysis},
  author={Kimoto, Koji and Kikkawa, Jun and Harano, Koji and Cretu, Ovidiu and Shibazaki, Yuki and Uesugi, Fumihiko},
  journal={Scientific Reports},
  volume={14},
  number={1},
  pages={2901},
  year={2024},
  publisher={Nature Publishing Group UK London}
}

@article{martis2023imaging,
  title={Imaging the electron charge density in monolayer MoS2 at the {\AA}ngstrom scale},
  author={Martis, Joel and Susarla, Sandhya and Rayabharam, Archith and Su, Cong and Paule, Timothy and Pelz, Philipp and Huff, Cassandra and Xu, Xintong and Li, Hao-Kun and Jaikissoon, Marc and others},
  journal={Nature communications},
  volume={14},
  number={1},
  pages={4363},
  year={2023},
  publisher={Nature Publishing Group UK London}
}

@article{nordahl2024exploring,
  title={Exploring deep learning models for 4D-STEM-DPC data processing},
  author={Nordahl, Gregory and Dagenborg, Sivert and S{\o}rhaug, J{\o}rgen and Nord, Magnus},
  journal={Ultramicroscopy},
  volume={267},
  pages={114058},
  year={2024},
  publisher={Elsevier}
}

@article{yuan2021machine,
  title={Machine Learning Based Precision Orientation and Strain Mapping from 4D Diffraction Datasets},
  author={Yuan, Renliang and Zhang, Jiong and He, Lingfeng and Zuo, Jian-Min},
  journal={Microscopy and Microanalysis},
  volume={27},
  number={S1},
  pages={1276--1278},
  year={2021},
  publisher={Cambridge University Press}
}

@article{bruefach2022analysis,
  title={Analysis of interpretable data representations for 4D-STEM using unsupervised learning},
  author={Bruefach, Alexandra and Ophus, Colin and Scott, Mary C},
  journal={Microscopy and Microanalysis},
  volume={28},
  number={6},
  pages={1998--2008},
  year={2022},
  publisher={Oxford University Press}
}

@inproceedings{ranieri2024assessing,
  title={Assessing the sensitivity of 4D-STEM measurements for electric field mapping at the sub-micrometer scale},
  author={Ranieri, Pierpaolo and Ignatans, Reinis and Boureau, Victor and Tileli, Vasiliki},
  booktitle={BIO Web of Conferences},
  volume={129},
  pages={04023},
  year={2024},
  organization={EDP Sciences}
}

@article{oxley2020deep,
  title={Deep learning of interface structures from simulated 4D STEM data: cation intermixing vs. roughening},
  author={Oxley, Mark P and Yin, Junqi and Borodinov, Nikolay and Somnath, Suhas and Ziatdinov, Maxim and Lupini, Andrew R and Jesse, Stephen and Vasudevan, Rama K and Kalinin, Sergei V},
  journal={Machine Learning: Science and Technology},
  volume={1},
  number={4},
  pages={04LT01},
  year={2020},
  publisher={IOP Publishing}
}

@article{bruefach2023robust,
  title={Robust design of semi-automated clustering models for 4D-STEM datasets},
  author={Bruefach, Alexandra and Ophus, Colin and Scott, MC},
  journal={APL Machine Learning},
  volume={1},
  number={1},
  year={2023},
  publisher={AIP Publishing}
}

@book{koch2002determination,
  title={Determination of core structure periodicity and point defect density along dislocations},
  author={Koch, Christoph Tobias},
  year={2002},
  publisher={Arizona State University}
}

@inproceedings{he2016deep,
  title={Deep residual learning for image recognition},
  author={He, Kaiming and Zhang, Xiangyu and Ren, Shaoqing and Sun, Jian},
  booktitle={Proceedings of the IEEE conference on computer vision and pattern recognition},
  pages={770--778},
  year={2016}
}

@article{simonyan2014very,
  title={Very deep convolutional networks for large-scale image recognition},
  author={Simonyan, Karen and Zisserman, Andrew},
  journal={arXiv preprint arXiv:1409.1556},
  year={2014}
}

@inproceedings{deng2009imagenet,
  title={Imagenet: A large-scale hierarchical image database},
  author={Deng, Jia and Dong, Wei and Socher, Richard and Li, Li-Jia and Li, Kai and Fei-Fei, Li},
  booktitle={2009 IEEE conference on computer vision and pattern recognition},
  pages={248--255},
  year={2009},
  organization={Ieee}
}

@article{Rojac_NM16,
	author = {Rojac, Tadej and Bencan, Andreja and Drazic, Goran and Sakamoto, Naonori and Ursic, Hana and Jancar, Bostjan and Tavcar, Gasper and Makarovic, Maja and Walker, Julian and Malic, Barbara and Damjanovic, Dragan},
	title = {Domain-wall conduction in ferroelectric {BiFeO$_3$} controlled by accumulation of charged defects},
	journal = {Nat. Mater.},
	volume = 16,
	pages = {322--327},
	year = 2017,
	month = mar,
	issn = {1476-4660},
	publisher = {Nature Publishing Group},
	doi = {10.1038/nmat4799}
}

@article{Bencan_NC12,
	author = {Bencan, Andreja and Oveisi, Emad and Hashemizadeh, Sina and Veerapandiyan, Vignaswaran K. and Hoshina, Takuya and Rojac, Tadej and Deluca, Marco and Drazic, Goran and Damjanovic, Dragan},
	title = {Atomic scale symmetry and polar nanoclusters in the paraelectric phase of ferroelectric materials},
	journal = {Nat. Commun.},
	volume = {12},
	number = {3509},
	pages = {1--9},
	year = {2021},
	month = jun,
	issn = {2041-1723},
	publisher = {Nature Publishing Group},
	doi = {10.1038/s41467-021-23600-3}
}

@article{Condurache_NL23,
	author = {Condurache, Oana and Dra{\ifmmode\check{z}\else\v{z}\fi}i{\ifmmode\acute{c}\else\'{c}\fi}, Goran and Rojac, Tadej and Ur{\ifmmode\check{s}\else\v{s}\fi}i{\ifmmode\check{c}\else\v{c}\fi}, Hana and Dkhil, Brahim and Brade{\ifmmode\check{s}\else\v{s}\fi}ko, Andra{\ifmmode\check{z}\else\v{z}\fi} and Damjanovic, Dragan and Ben{\ifmmode\check{c}\else\v{c}\fi}an, Andreja},
	title = {Atomic-Level Response of the Domain Walls in Bismuth Ferrite in a Subcoercive-Field Regime},
	journal = {Nano Lett.},
	volume = {23},
	number = {2},
	pages = {750--756},
	year = {2023},
	month = jan,
	issn = {1530-6984},
	publisher = {American Chemical Society},
	doi = {10.1021/acs.nanolett.2c02857}
}

@article{Chen_NC13,
	author = {Chen, Liang and Deng, Shiqing and Liu, Hui and Wu, Jie and Qi, He and Chen, Jun},
	title = {Giant energy-storage density with ultrahigh efficiency in lead-free relaxors via high-entropy design},
	journal = {Nat. Commun.},
	volume = {13},
	number = {3089},
	pages = {1--8},
	year = {2022},
	month = jun,
	issn = {2041-1723},
	publisher = {Nature Publishing Group},
	doi = {10.1038/s41467-022-30821-7}
}

@article{ludacka2024imaging,
  title={Imaging and structure analysis of ferroelectric domains, domain walls, and vortices by scanning electron diffraction},
  author={Ludacka, Ursula and He, Jiali and Qin, Shuyu and Zahn, Manuel and Christiansen, Emil Frang and Hunnestad, Kasper A and Zhang, Xinqiao and Yan, Zewu and Bourret, Edith and K{\'e}zsm{\'a}rki, Istv{\'a}n and others},
  journal={npj computational materials},
  volume={10},
  number={1},
  pages={106},
  year={2024},
  publisher={Nature Publishing Group UK London}
}

@article{hardy2025polarization,
  title={Polarization Domain Mapping From 4D-STEM Using Deep Learning},
  author={Hardy, Fintan G and Griffin, Sinead M and Palos, Mariana and Li, Yaqi and Topore, Geri and Walsh, Aron and Conroy, Michele Shelly},
  journal={arXiv preprint arXiv:2510.00693},
  year={2025}
}

@article{condurache2021atomically,
  title={Atomically resolved structure of step-like uncharged and charged domain walls in polycrystalline BiFeO3},
  author={Condurache, Oana and Dra{\v{z}}i{\'c}, Goran and Sakamoto, Naonori and Rojac, Tadej and Ben{\v{c}}an, Andreja},
  journal={Journal of Applied Physics},
  volume={129},
  number={5},
  year={2021},
  publisher={AIP Publishing}
}

@article{zhang2021lead,
  title={Lead-free ferroelectric materials: Prospective applications},
  author={Zhang, Shujun and Mali{\v{c}}, Barbara and Li, Jing-Feng and R{\"o}del, J{\"u}rgen},
  journal={Journal of Materials Research},
  volume={36},
  number={5},
  pages={985--995},
  year={2021},
  publisher={Springer}
}

@article{Giannozzi_JPCM21,
  author={Paolo Giannozzi and Stefano Baroni and Nicola Bonini and
                  Matteo Calandra and Roberto Car and Carlo Cavazzoni
                  and Davide Ceresoli and Guido L Chiarotti and Matteo
                  Cococcioni and Ismaila Dabo and Dal Corso, Andrea  and
                  Stefano de Gironcoli and Stefano Fabris and Guido
                  Fratesi and Ralph Gebauer and Uwe Gerstmann and
                  Christos Gougoussis and Anton Kokalj and Michele
                  Lazzeri and Layla Martin-Samos and Nicola Marzari
                  and Francesco Mauri and Riccardo Mazzarello and
                  Stefano Paolini and Alfredo Pasquarello and Lorenzo
                  Paulatto and Carlo Sbraccia and Sandro Scandolo and
                  Gabriele Sclauzero and Ari P Seitsonen and Alexander
                  Smogunov and Paolo Umari and Renata M Wentzcovitch},
  title={\textsc{Quantum} {ESPRESSO}: a modular and open-source software project for quantum simulations of materials},
  journal={J. Phys: Condens. Matter},
  volume={21},
  number={39},
  pages={395502},
  doi = {10.1088/0953-8984/21/39/395502},
  url = {https://iopscience.iop.org/article/10.1088/0953-8984/21/39/395502},
  year={2009},
  note = {Code available from \url{http://www.quantum-espresso.org/}}
}

@article{Giannozzi_JPCM29,
  author={Paolo Giannozzi and Oliviero Andreussi and Thomas Brumme and
                  Oana Bunau and Marco Buongiorno Nardelli and Matteo
                  Calandra and Roberto Car and Carlo Cavazzoni and
                  Davide Ceresoli and Matteo Cococcioni and Nicola
                  Colonna and Ivan Carnimeo and Andrea Dal Corso and
                  Stefano de Gironcoli and Pietro Delugas and Robert
                  DiStasio and Andrea Ferretti and Andrea Floris and
                  Guido Fratesi and Giorgia Fugallo and Ralph Gebauer
                  and Uwe Gerstmann and Feliciano Giustino and Tommaso
                  Gorni and Junteng Jia and Mitsuaki Kawamura and
                  Hsin-Yu Ko and Anton Kokalj and Emine Küçükbenli and
                  Michele Lazzeri and Margherita Marsili and Nicola
                  Marzari and Francesco Mauri and Ngoc Linh Nguyen and
                  Huy-Viet Nguyen and Alberto Otero-de-la-Roza and
                  Lorenzo Paulatto and Samuel Poncé and Dario Rocca
                  and Riccardo Sabatini and Biswajit Santra and Martin
                  Schlipf and Ari Paavo Seitsonen and Alexander
                  Smogunov and Iurii Timrov and Timo Thonhauser and
                  Paolo Umari and Nathalie Vast and Xifan Wu and
                  Stefano Baroni},
  title={Advanced capabilities for materials modelling with \textsc{Quantum} {ESPRESSO}},
  journal={J. Phys: Condens. Matter},        
  url={http://iopscience.iop.org/10.1088/1361-648X/aa8f79},
  doi={10.1088/1361-648X/aa8f79},
  year={2017},
  volume={29},
  pages={465901}
}

@Article{Perdew_PRL77,
  title = {Generalized gradient approximation made simple},
  author = {Perdew, John P. and Burke, Kieron  and Ernzerhof, Matthias },
  journal = {Phys. Rev. Lett.},
  volume = {77},
  number = {18},
  pages = {3865--3868},
  numpages = {3},
  year = {1996},
  month = {Oct},
  doi = {10.1103/PhysRevLett.77.3865},
  url = {https://journals.aps.org/prl/abstract/10.1103/PhysRevLett.77.3865},
  publisher = {American Physical Society}
}

@article{DalCorso_CMS95,
title = "Pseudopotentials periodic table: {F}rom {H} to {P}u",
journal = "Comp. Mater. Sci.",
volume = "95",
pages = "337--350",
year = "2014",
OPTissn = "0927-0256",
doi = "10.1016/j.commatsci.2014.07.043",
url = "http://www.sciencedirect.com/science/article/pii/S0927025614005187",
author = "Dal Corso, Andrea",
}

@Article{Blochl_PRB50,
  title = {Projector augmented-wave method},
  author = {Bl\"ochl, P. E.},
  journal = {Phys. Rev. B},
  volume = {50},
  number = {24},
  pages = {17953--17979},
  numpages = {26},
  year = {1994},
  month = {Dec},
  doi = {10.1103/PhysRevB.50.17953},
  publisher = {American Physical Society}
}

@Misc{PAW,
note = {{PAW} pseudopotentials for the O, Na, K, and Nb atoms
                  were taken from the pslibrary
                  (files: {\tt O.pbe-n-kjpaw\_psl.1.0.0.UPF},
                  {\tt Na.pbe-spn-kjpaw\_psl.0.2.UPF},
                  {\tt K.pbe-spn-kjpaw\_psl.1.0.0.UPF},
                  {\tt Nb.pbe-spn-kjpaw\_psl.0.3.0.UPF})},
}

@article{Anisimov_PRB44,
  title = {Band theory and {M}ott insulators: {H}ubbard {U} instead of {S}toner {I}},
  author = {Anisimov, Vladimir I. and Zaanen, Jan and Andersen, Ole K.},
  journal = {Phys. Rev. B},
  volume = {44},
  issue = {3},
  pages = {943--954},
  year = {1991},
  month = {Jul},
  publisher = {American Physical Society},
  doi = {10.1103/PhysRevB.44.943},
  url = {https://link.aps.org/doi/10.1103/PhysRevB.44.943}
}

@article{Cococcioni_PRB71,
  title = {Linear response approach to the calculation of the effective interaction parameters in the $\mathrm{LDA}+\mathrm{U}$ method},
  author = {Cococcioni, Matteo and de Gironcoli, Stefano},
  journal = {Phys. Rev. B},
  volume = {71},
  issue = {3},
  pages = {035105},
  numpages = {16},
  year = {2005},
  month = {Jan},
  publisher = {American Physical Society},
  doi = {10.1103/PhysRevB.71.035105},
  url = {https://link.aps.org/doi/10.1103/PhysRevB.71.035105}
}

@article{Timrov_PRB98,
        author = {Timrov, Iurii and Marzari, Nicola and Cococcioni, Matteo},
        title = {Hubbard parameters from density-functional perturbation theory},
        journal = {Phys. Rev. B},
        volume = {98},
        number = {8},
        pages = {085127},
        year = {2018},
        month = aug,
        issn = {2469-9969},
        publisher = {American Physical Society},
        doi = {10.1103/PhysRevB.98.085127}
}

\section{Acknowledgments}

This work is supported by the Slovenian Research and Innovation Agency (Project No.\@{} P2-0105, P2-0103, J7-4637, and GC-0001).

\section{Author information}

\subsection{Authors and Affiliations}

\textbf{Jožef Stefan Institute, Jamova 39, Ljubljana, Slovenia}\\ 
Matej Martinc, Goran Dražič, Anton Kokalj, Katarina Žiberna, Janina Roknić, Matic Poberžnik, Sašo Džeroski \& Andreja Benčan Golob

\subsection{Contributions} 

M.M., A.B.G. and G.D. conceived the study and designed the experimental and computational framework. M.M. developed the machine learning models, performed the benchmarking, and implemented the prototype representation training and PCA-based workflows. G.D., A.B.G., K.Ž and J.R. performed the 4D-STEM experiments and HAADF-STEM imaging and image simulations. A.K. and M.P. provided expertise in DFT calculations. K.Ž. and J.R. contributed to the manual labeling of experimental 4D-STEM datasets. S.D. supervised the machine learning methodology and provided critical insights into the evaluation metrics. A.B.G. supervised the overall research project. M.M., G.D. and A.B.G. drafted the manuscript. All authors contributed to the discussions and provided feedback on the final manuscript.

\section{Ethics declarations}

\subsection{Competing interests} 

The authors declare no competing interests.

\end{document}